\documentstyle[jkas]{article}

\beginpage{1}
\endpage{9}
\year{2012}\volume{00}\month{October}

\runningauthor {J.-H. PARK \& S. TRIPPE} 
\runningtitle{MULTIPLE EMISSION STATES IN AGN} 

\month{October} \year{2012} \volume{00} \issueno{0}
\beginpage{1}\endpage{9}
\date{Received 2012 September 17; Revised 2012 October 20; Accepted 2012 October 26}

\begin{document}
\title{MULTIPLE EMISSION STATES IN ACTIVE GALACTIC NUCLEI} 

\author{Jong-Ho Park \& Sascha Trippe\vspace{0.25mm}} 
\address{Department of Physics and Astronomy, Seoul National University,
Seoul 151-742, South Korea\\ {\it E-mail: goldtank90@snu.ac.kr; trippe@astro.snu.ac.kr}}

\address{\normalsize{\it (Received 2012 September 17; Revised 2012 October 20; Accepted 2012 October 26)}}
\offprints{S. Trippe}

\def\thesection{\Roman{section}}
\def\thesubsection{\Alph{subsection}.}
\def\thesubsubsection{\arabic{subsubsection}.}

\abstract{\noindent We present a test of the emission statistics of active galactic nuclei (AGN), probing the connection between red-noise temporal power spectra and multi-modal flux distributions known from observations. We simulate AGN lightcurves under the assumption of uniform stochastic emission processes for different power-law indices of their power spectra. For sufficiently shallow slopes (power-law indices $\beta\lesssim1$), the flux distributions (histograms) of the resulting lightcurves are approximately Gaussian. For indices corresponding to steeper slopes ($\beta\gtrsim1$), the flux distributions become multi-modal. This finding disagrees systematically with results of recent mm/radio observations. Accordingly, we conclude that the emission from AGN does not necessarily originate from uniform stochastic processes even if their power spectra suggest this. Possible mechanisms are transitions between different activity states and/or the presence of multiple, spatially disconnected, emission regions.}

\maketitle


\section{INTRODUCTION}

\noindent
There is general consensus that active galactic nuclei (AGN) are powered by accretion of matter onto supermassive ($M_{\bullet}\approx10^{6...10}\,M_{\odot}$) black holes located in the centers of most, if not all, galaxies (see e.g. \citealt{pet}, \citealt{krol}, or \citealt{beckmann2012}, for reviews). AGN are known for strong temporal variability of their luminosity. Especially at radio frequencies, variability appears to be related to shocks propagating through jets (e.g. \citealt{marscher}). At optical, ultraviolet, and higher frequencies changes in accretion flows and accretion disk structure are supposed to further modulate the luminosity (e.g. \citealt{devries2005,czerny}).

To understand the underlying physics of the variability, several studies aimed at identifying characteristic variability timescales. Timescales probed by observations range from few kiloseconds (\citealt{ben}; using X-ray observations, \citealt{schodel}; using radio observations) to tens of years (\citealt{hov7}, 2008; using radio observations). Other studies analyzed the temporal power spectra -- the square moduli of the Fourier transforms -- of AGN lightcurves. \citet{pres} pointed out that power spectra of AGN lightcurves follow power laws $A_f{\propto}f^{-\beta}$ with $\beta>0$, i.e. \emph{red noise} laws; here $A_f$ denotes the power spectral amplitude for a given sampling frequency $f$.\footnote{According to this convention, a power-law curve with index $\beta$ has a slope of $-\beta$.} Cases commonly encountered in statistics are \emph{white noise} ($\beta = 0$), \emph{random walk noise} ($\beta = 2$), and (intermediate) \emph{flicker noise} ($\beta = 1$). In many cases observed, AGN power spectra can be described by $1/f$ flicker noise \citep{law1987}. Red-noise power spectra are analyzed in depth by \cite{law} who used 12 high-quality ``long look'' X-ray lightcurves of AGN from the EXOSAT data base, as well as by \cite{utl} who found observational evidence for a flattening of AGN X-ray power spectra at timescales longer than few months. At optical wavelengths, \citet{devries2005} found evidence for the variability of quasar emission following red-noise laws on timescales as long as approximately 40 years.

Whether the emission processes of AGN are indeed uniform stochastic has been debated hotly especially for the case of Sagittarius~A*, the supermassive black hole located at the center of the Milky Way. Some studies argued that power spectra of its near-infrared lightcurves are consistent with a single red noise law \citep{Mey8,do2009}. Other studies found indication that there are multiple states of emission, possibly including a quasi-periodic modulation of the source flux (\citealt{gil}, \citealt{trip7}, \citealt{dod}).

Recent millimeter/radio observations \citep{trip} indicate that the emission statistics of AGN is more complex than generally assumed. Whereas the lightcurves of the sample of \citet{trip} showed the known red noise power spectra, the flux distributions (histograms) showed clear bi- or multi-modality. This result was somewhat unexpected as power spectra with $\beta<1$, as observed by \citet{trip}, can be understood as being generated by lightcurves with flux values drawn from uni-modal -- approximately Gaussian -- distributions. The observations were interpreted qualitatively as indication for the presence of distinct emission states.

A \emph{quantitative} test of this interpretation requires a careful evaluation of the relation between the slope of the power spectrum and the corresponding flux distribution. Red noise power spectra imply higher variability at longer timescales, and realistic lightcurves as well as the corresponding power spectra are noisy. Accordingly, extreme flux values or multi-modal flux distributions may occur under certain conditions even if not expected in general.

\section{ANALYSIS}

\begin{figure*}[!t]
\centering
\epsfxsize = 5.7cm \epsfbox{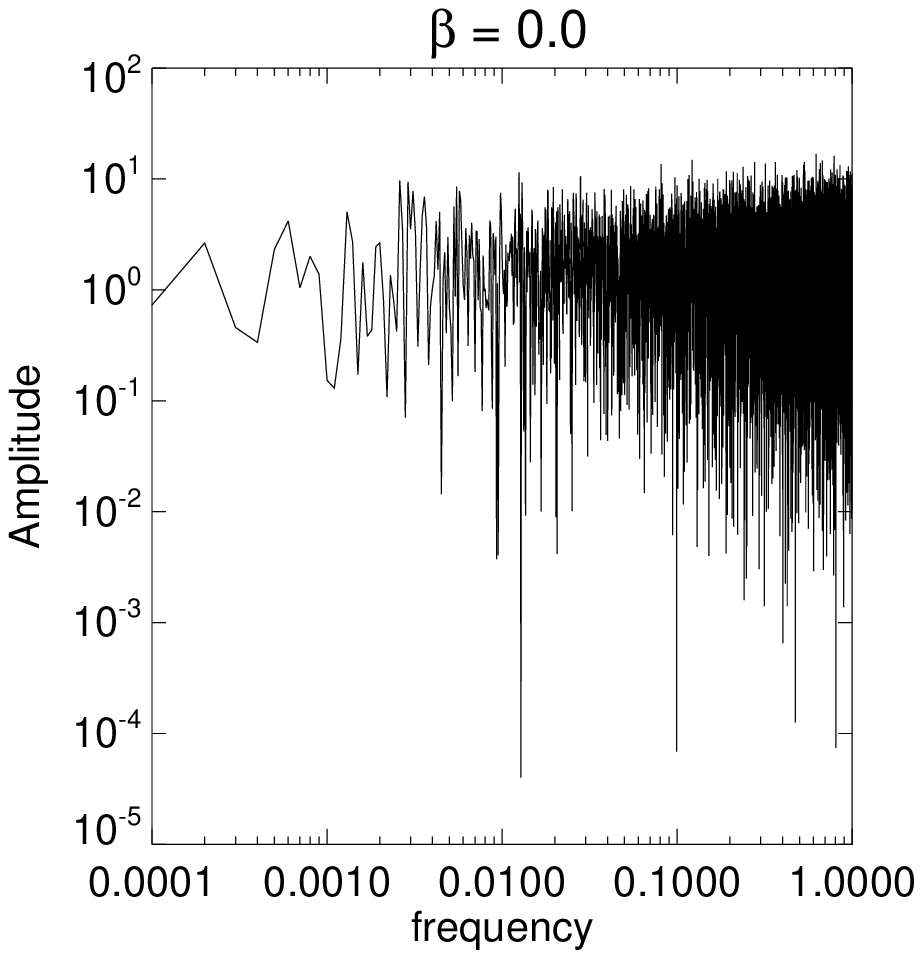}
\epsfxsize = 5.7cm \epsfbox{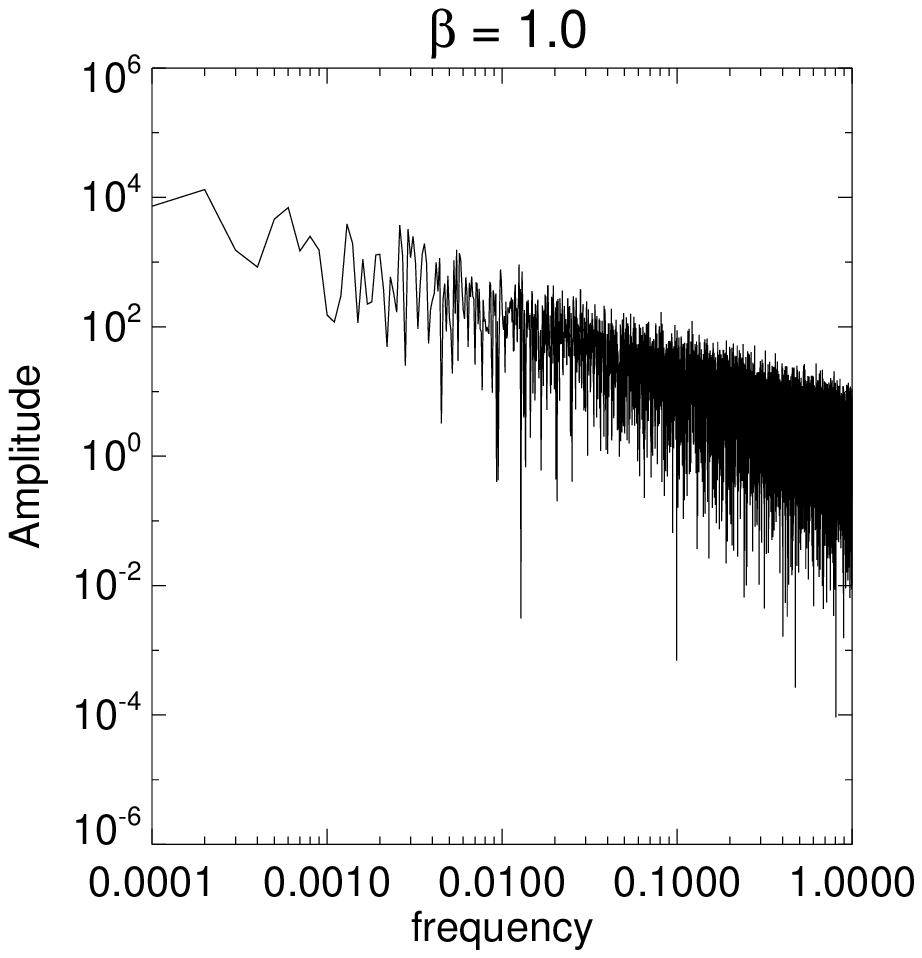}
\epsfxsize = 5.7cm \epsfbox{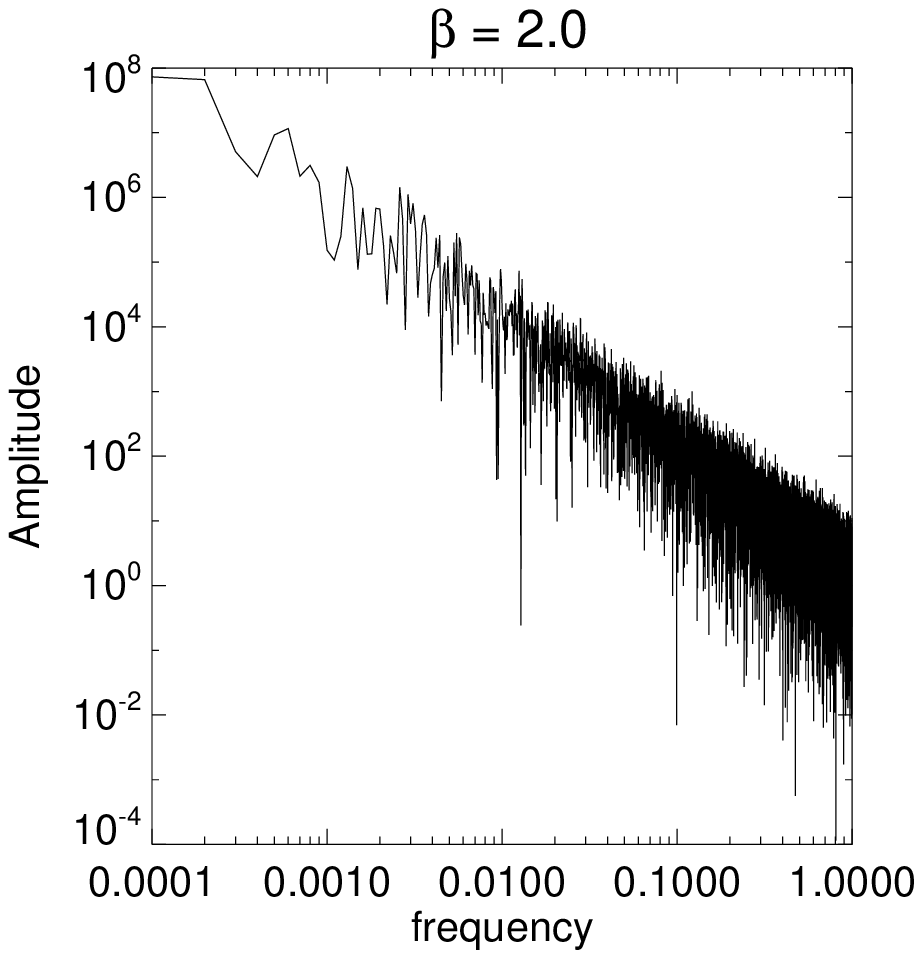}
\caption{Simulated power spectra with $\beta = 0, 1, 2$ (from left to right) in logarithmic scales. Units are arbitrary. These artificial power spectra agree with actual observations (compare, e.g. \citealt{tim}, or \citealt{trip}).}
\label{powerspectra}
\end{figure*}

\subsection{Simulated Lightcurves}

\noindent
Monte Carlo simulations provide a convenient way to probe the relations between red noise power spectra and the underlying flux distributions. We generate artificial Fourier transforms of lightcurves with $\beta = 0$, 0.25, 0.5, 0.75, 1, 1.25, 1.5, 1.75, and 2 using the algorithm introduced by \cite{tim}. For each sampling frequency $f$, we draw two random numbers from Gaussian distributions for the real part and the imaginary part, respectively. Power-law noise with slope $-\beta$ is generated by multiplying both values with $f^{-\beta/2}$. The individual values are stored in an array of complex numbers; this array corresponds to the complex Fourier transform of our artificial lightcurve. We calculate the normalized temporal power spectrum, also known as \emph{periodogram}, by taking the square of the absolute value of the array. A periodogram is given by

\begin{equation}
\footnotesize
\left|F(\omega)\right|^2=\frac{1}{N}{\displaystyle\biggl[\Big(\sum_{t}S(t)\,{\cos}\,{\omega}t\Big)^2+\Big(\sum_{t}S(t)\,{\sin}\,{\omega}t\Big)^2\biggr]}
\end{equation}

{\noindent}\citep{sch} where $S(t)$ is the signal in time domain, $F(\omega)$ is the Fourier transform, and $\omega=2{\pi}f$. By construction, a periodogram of our simulated signal follows the relation $A_f{\propto}f^{-\beta}$. The results agree with power spectra of actual observations (see Fig. \ref{powerspectra}).

We obtain simulated lightcurves $S(t)$ by taking the inverse Fourier transform of our complex arrays. Here we use the inverse Fourier transform of the FFT (Fast Fourier Transform) procedure in IDL.\footnote{Interactive Data Language, ITT Exelis Inc., McLean (Virginia).} The inverse Fourier transform is given by

\begin{equation}
S(t) = \sum_{f=-N/2}^{N/2}F(f)\,{\exp}[i 2{\pi}ft/N]
\end{equation}

\noindent
where $t$ is the time, $f$ denotes the sampling frequency, $F(f)$ are the values in frequency domain and $N$ is the total number of array elements. To obtain a real valued time series, we construct our complex arrays such that $F(-f_i) = F^{\ast}(f_i)$, where the operator $^{\ast}$ denotes complex conjugation \citep{tim}. For each value of $\beta$ we simulate $N'$ -- meaning the number of trials -- lightcurves.

Each artificial lightcurve is composed of 20\,000 data points initially. However, in order to circumvent intrinsic symmetries of the Fourier transform process, it is necessary to restrict the analysis to a subset of -- adjacent -- data points taken from one half of the time series. A priori, the choice of a certain number of data points should not influence the properties of the resulting flux distributions; this is due to the use of power-law noise and the fact that power-laws are scale free. Nevertheless, to minimize the risk of selection effects, we perform two separate analysis runs:

\begin{itemize}

\item[$\triangleright$]  {\sc Run A} uses $N=9\,000$ data points per time series, the number of trials is $N'=100$.

\item[$\triangleright$]  {\sc Run B} uses $N=2\,000$ data points per time series, the number of trials is $N'=500$.

\end{itemize}

In order to ease comparison among different realizations of lightcurves, we use the normalization

\begin{equation} \label{norm}
S_{\rm norm}(t_i) = \frac{S(t_i) - \overline {S(t)}}{\sigma}
\end{equation}

\noindent
where $S(t)$ is the lightcurve, $\overline{S(t)}$ denotes the mean value of $S(t)$, and $\sigma$ is the standard deviation of $S(t)$. Accordingly, our time series are normalized to zero mean and unity standard deviation.

We check that our choices of $N$ do not alter the values of $\beta$ -- which were applied to time series with 20\,000 data points initially -- systematically. For this check, we obtain the actual $\beta$ values by calculating for each lightcurve the normalized \emph{Scargle periodogram}

\begin{equation}
\footnotesize
A_f=\frac{1}{2\sigma^2}{\left[ \frac{\Big({\sum_{i}{S_i}\,{{\cos}}\,2{\pi}ft_i}\Big)^2}{\sum_{i}{{{\cos}^2}2{\pi}ft_i}}+\frac{\Big({\sum_{i}{S_i}\,{{\sin}}\,2{\pi}ft_i}\Big)^2}{\sum_{i}{{{\sin}^2}2{\pi}ft_i}} \right]}
\end{equation}

\noindent
\citep{scar}. Here $A_f$ is the amplitude of the periodogram evaluated at frequency $f$, which is equivalent to the power spectral density;\footnote{For the analysis of observational data, the Scargle periodogram is preferable over the standard periodogram as it can be applied to data with arbitrary sampling and has a well-understood statistical behavior.} $S_i$ are the flux value and $t_i$ the time of the $i$th data point, respectively; and $\sigma^2$ is the variance of the data. The base frequency is $f_{\rm min}$; the $i$th frequency is $i$ times of $f_{\rm min}$, i.e. $f=f_{\rm min},2f_{\rm min},3f_{\rm min},\ldots,f_{\rm max}$ with $f_{\rm min} = 1/T, f_{\rm max}=N/(2T)$. Here $T$ is the total observation time and $N$ is the number of data points (9\,000 for {\sc run A} and 2\,000 for {\sc run B}). We confirm that input values and actual values of $\beta$ agree within $\approx$ 2\%.

\subsection{Exploring the Flux Distributions}

\begin{figure*}
\centering
\epsfxsize = 4.25cm \epsfbox{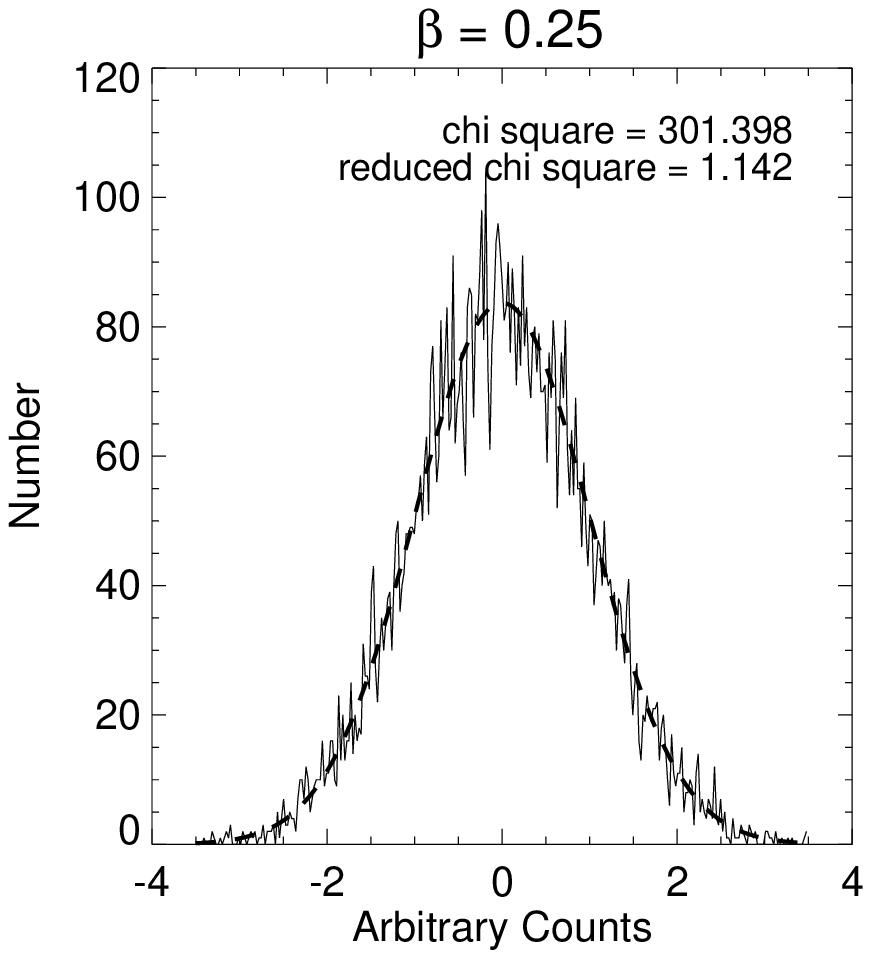}
\epsfxsize = 4.25cm \epsfbox{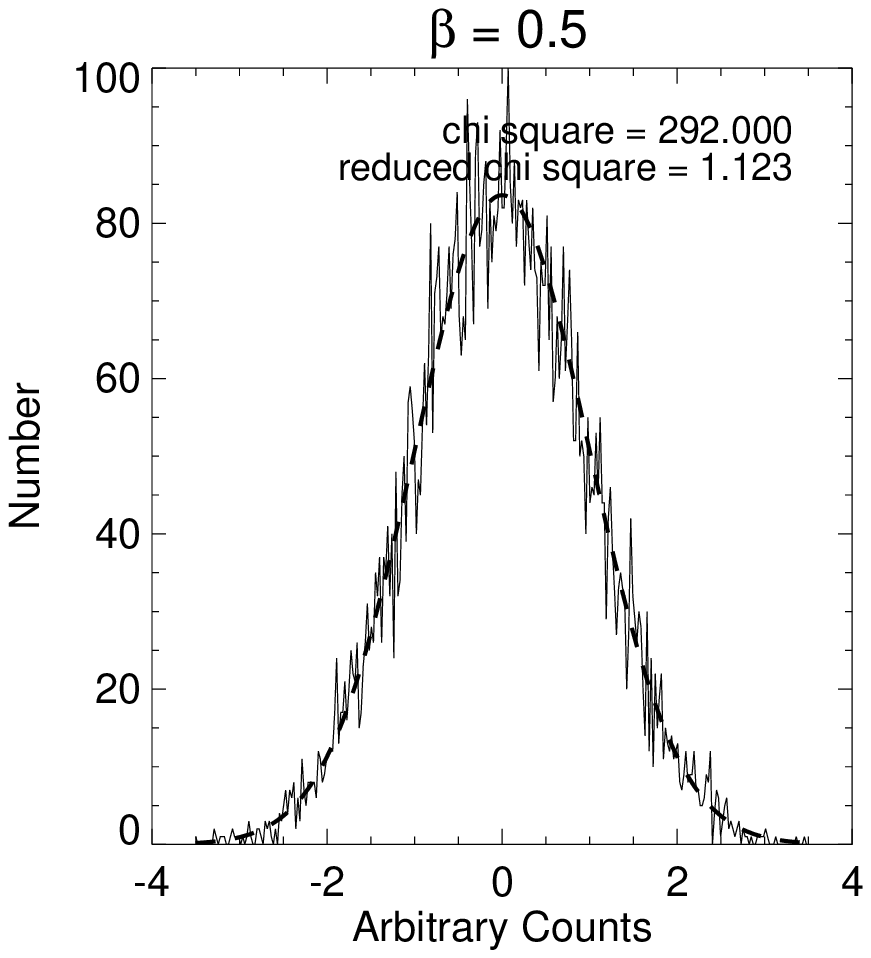}
\epsfxsize = 4.25cm \epsfbox{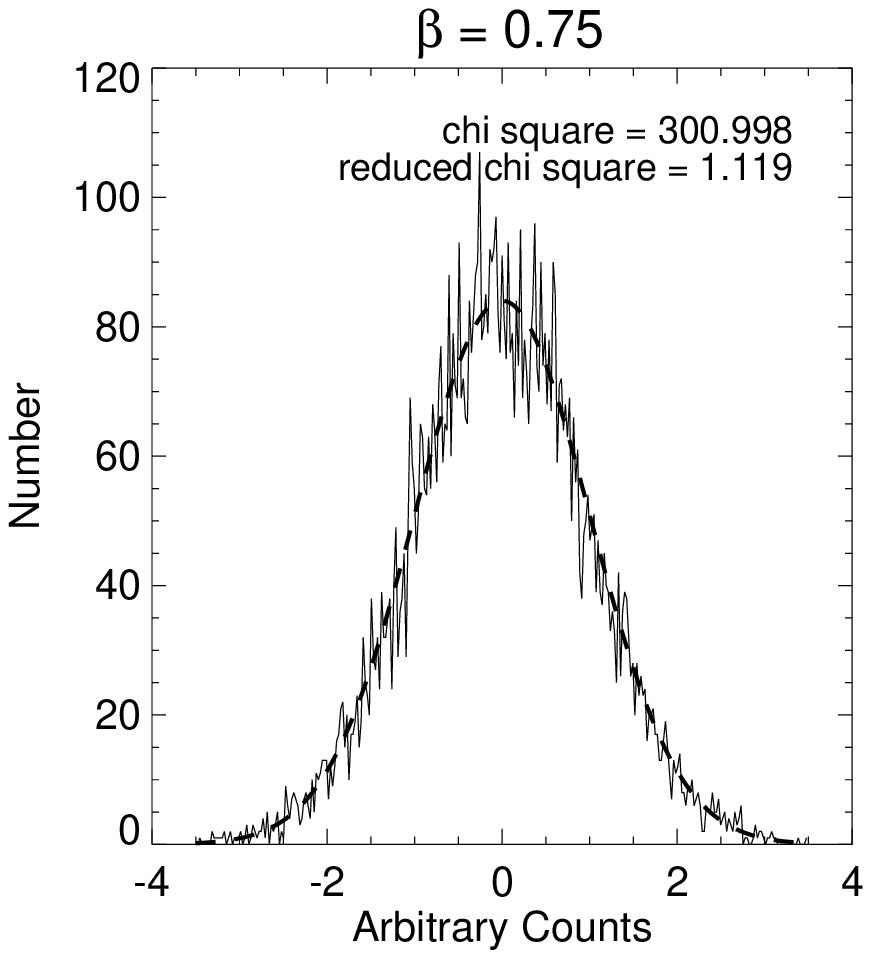}
\epsfxsize = 4.25cm \epsfbox{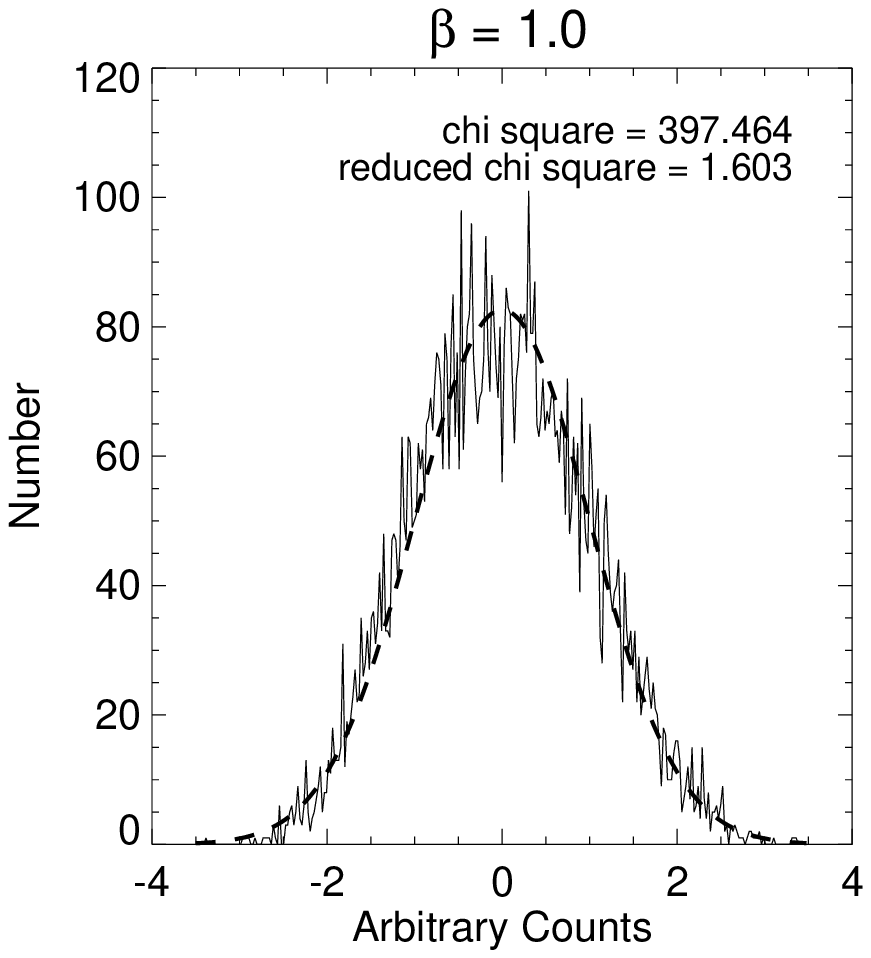} \\
\vspace{2mm}
\epsfxsize = 4.25cm \epsfbox{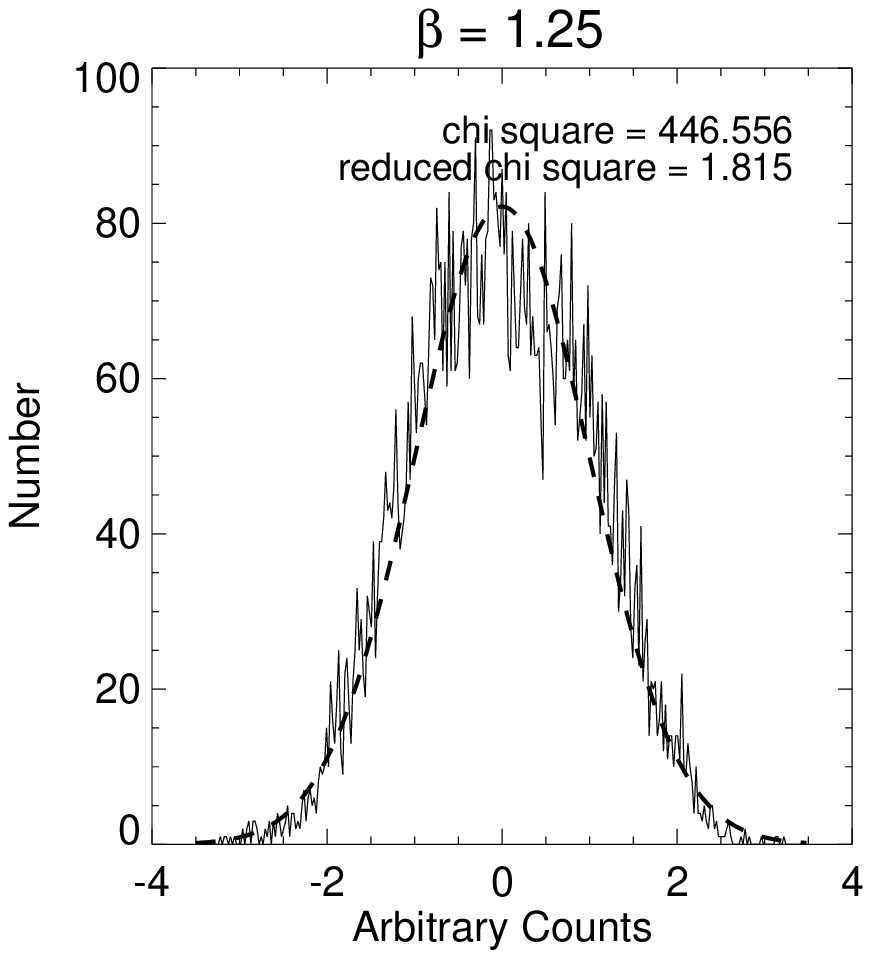}
\epsfxsize = 4.25cm \epsfbox{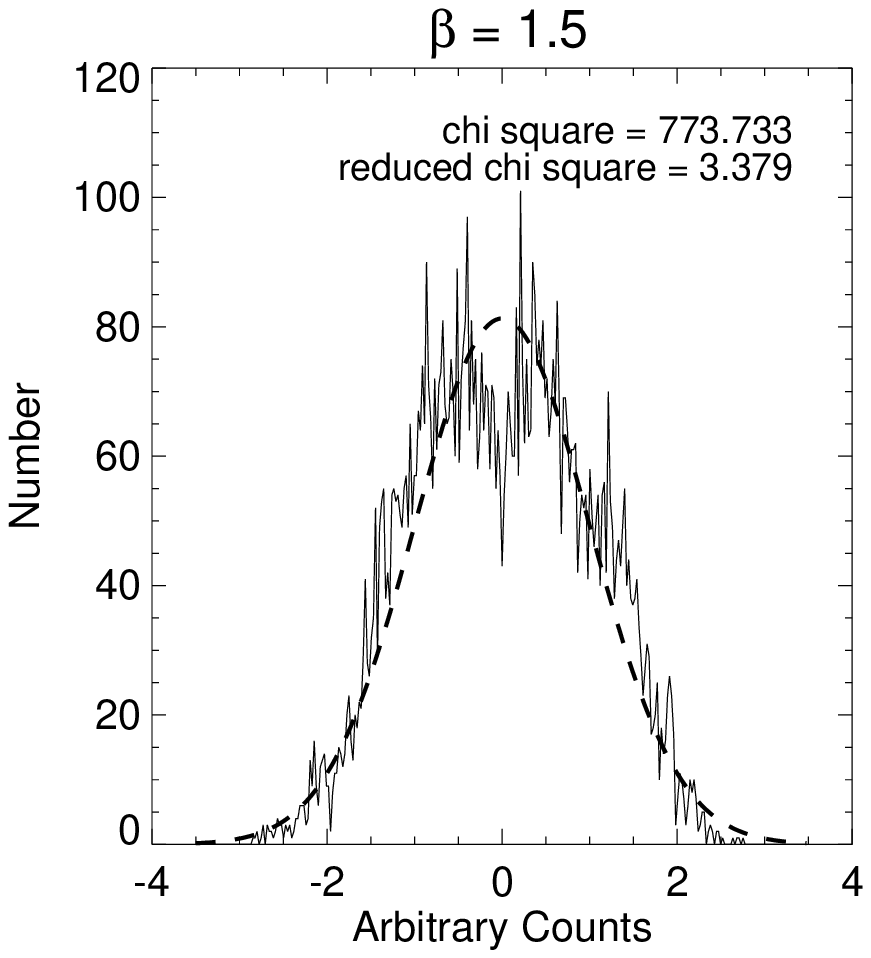}
\epsfxsize = 4.25cm \epsfbox{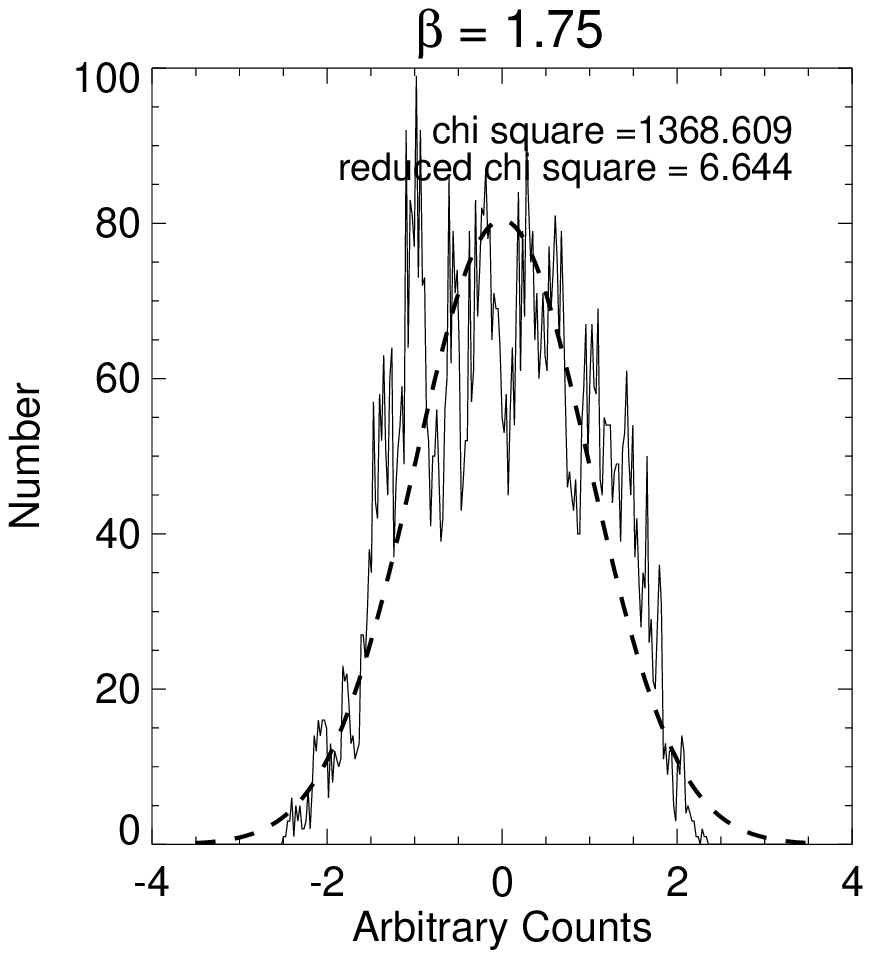}
\epsfxsize = 4.25cm \epsfbox{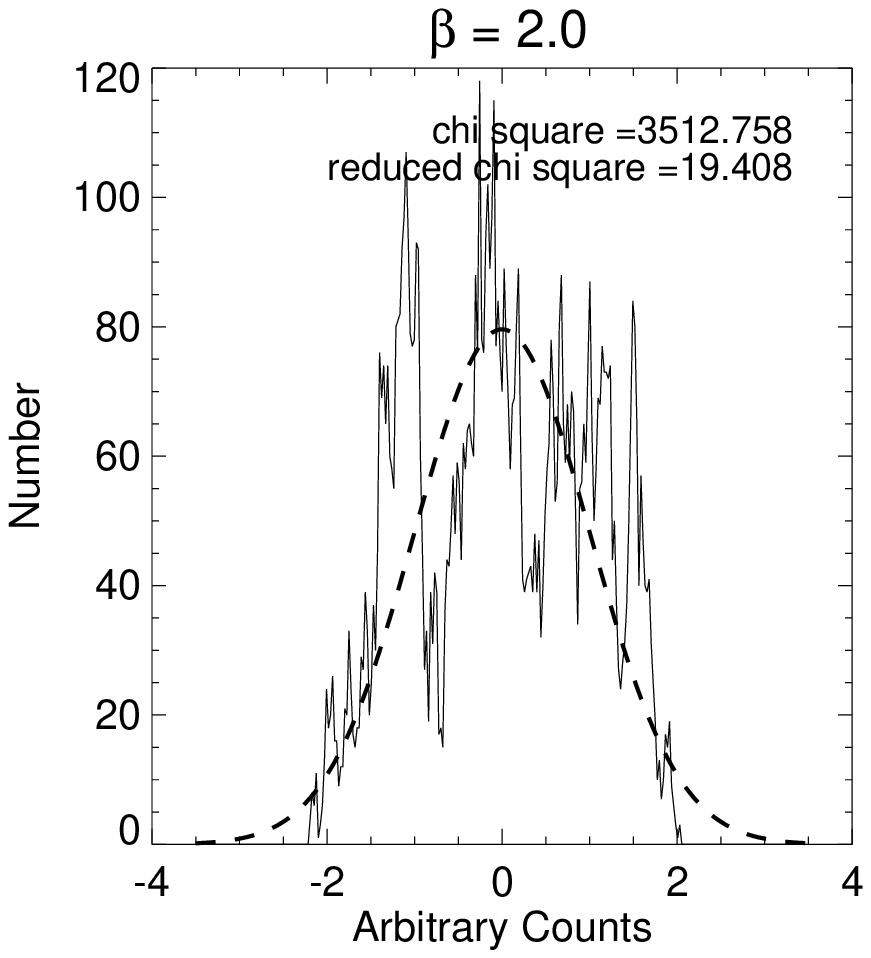} \\
\hrulefill \\
\vspace{6mm}
\epsfxsize = 4.25cm \epsfbox{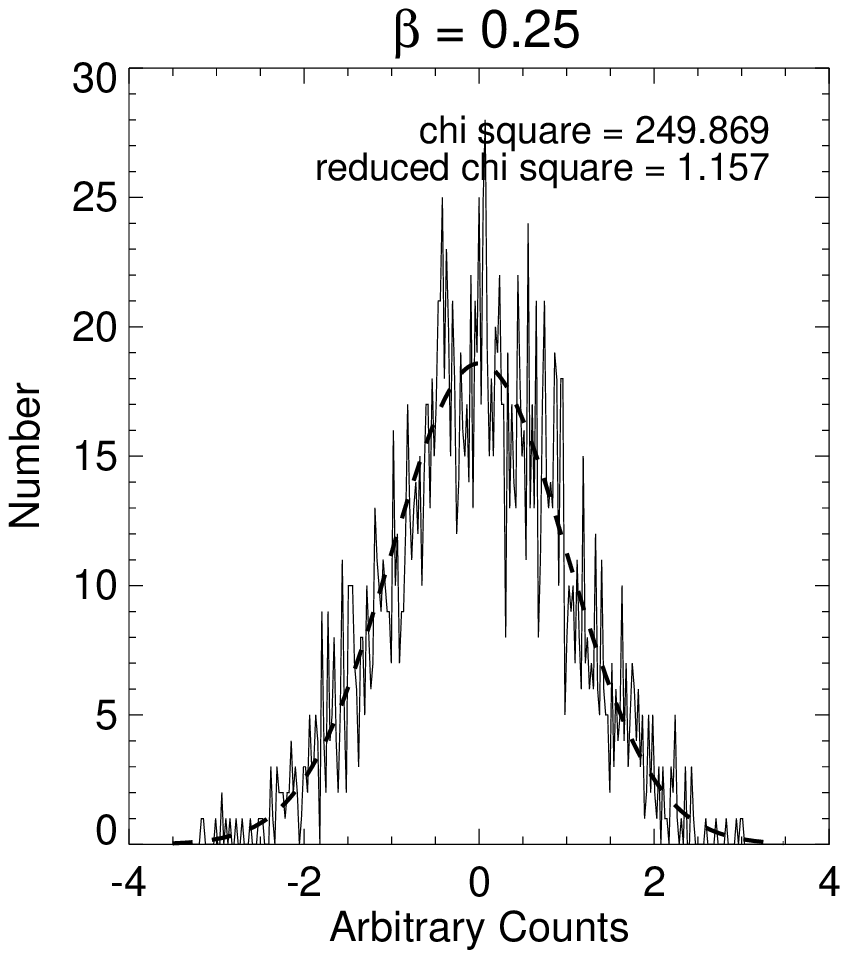}
\epsfxsize = 4.25cm \epsfbox{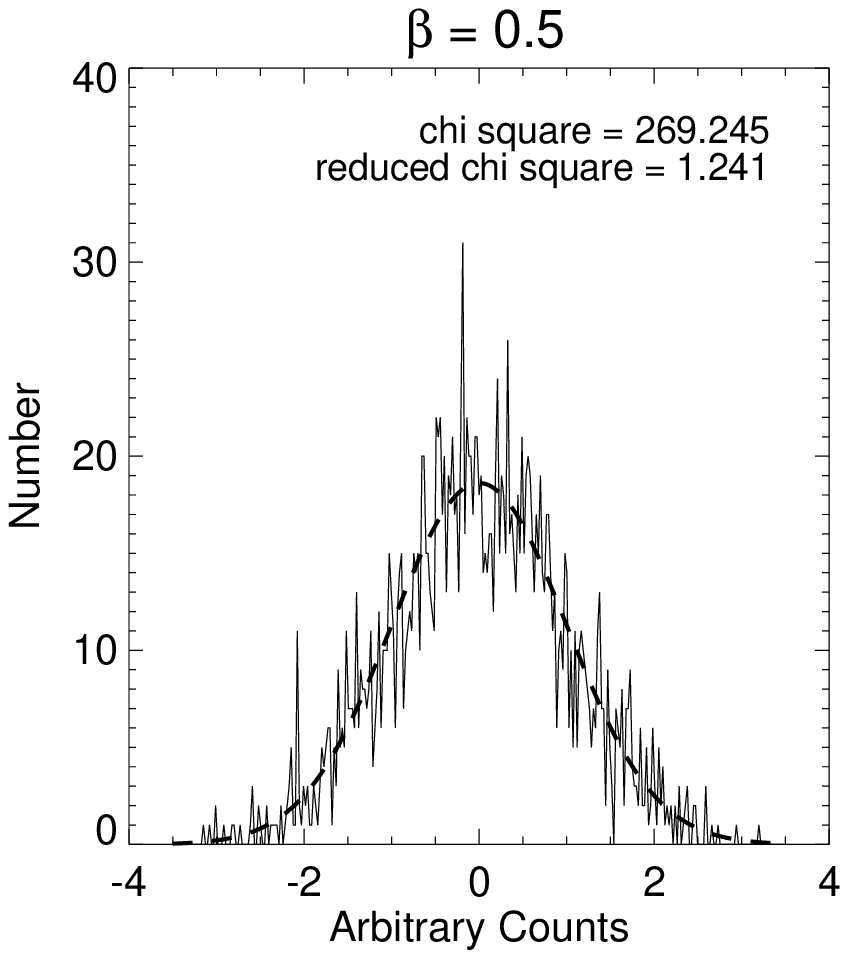}
\epsfxsize = 4.25cm \epsfbox{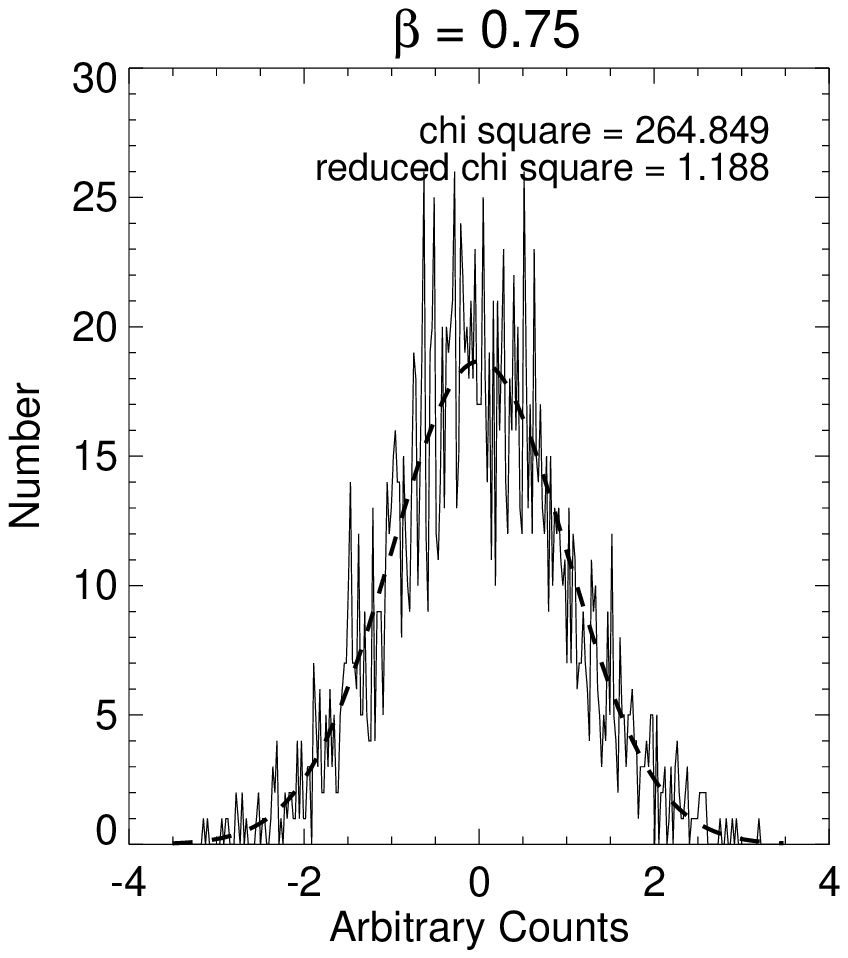}
\epsfxsize = 4.25cm \epsfbox{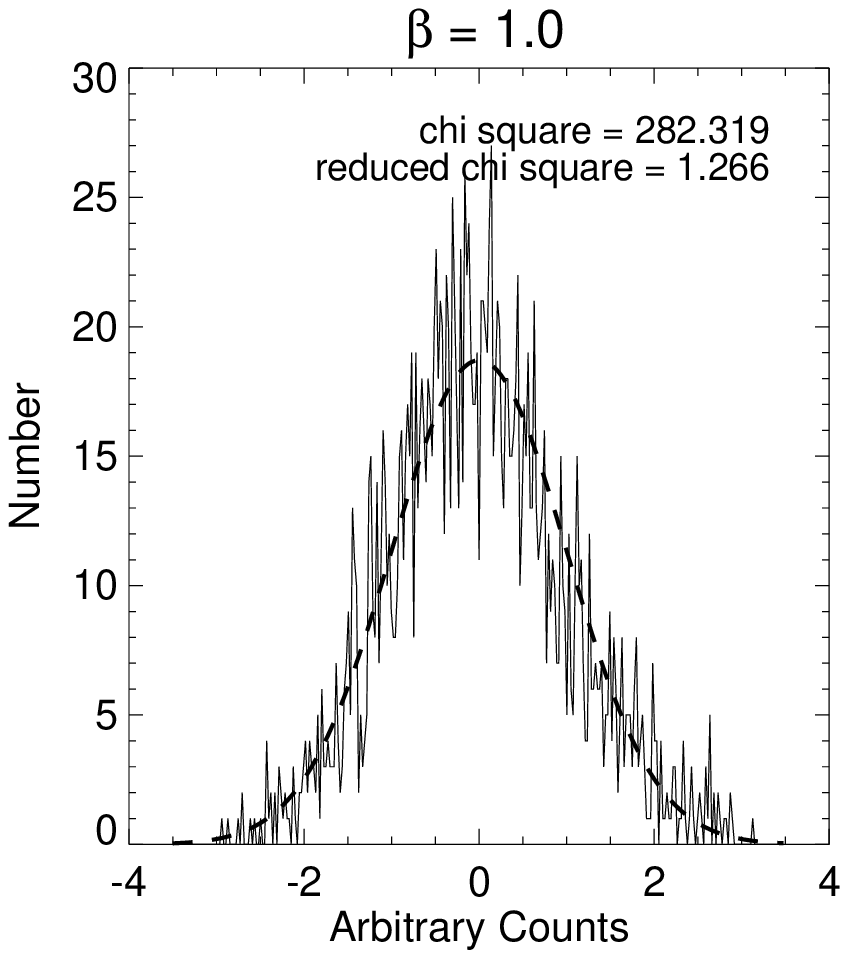} \\
\vspace{2mm}
\epsfxsize = 4.25cm \epsfbox{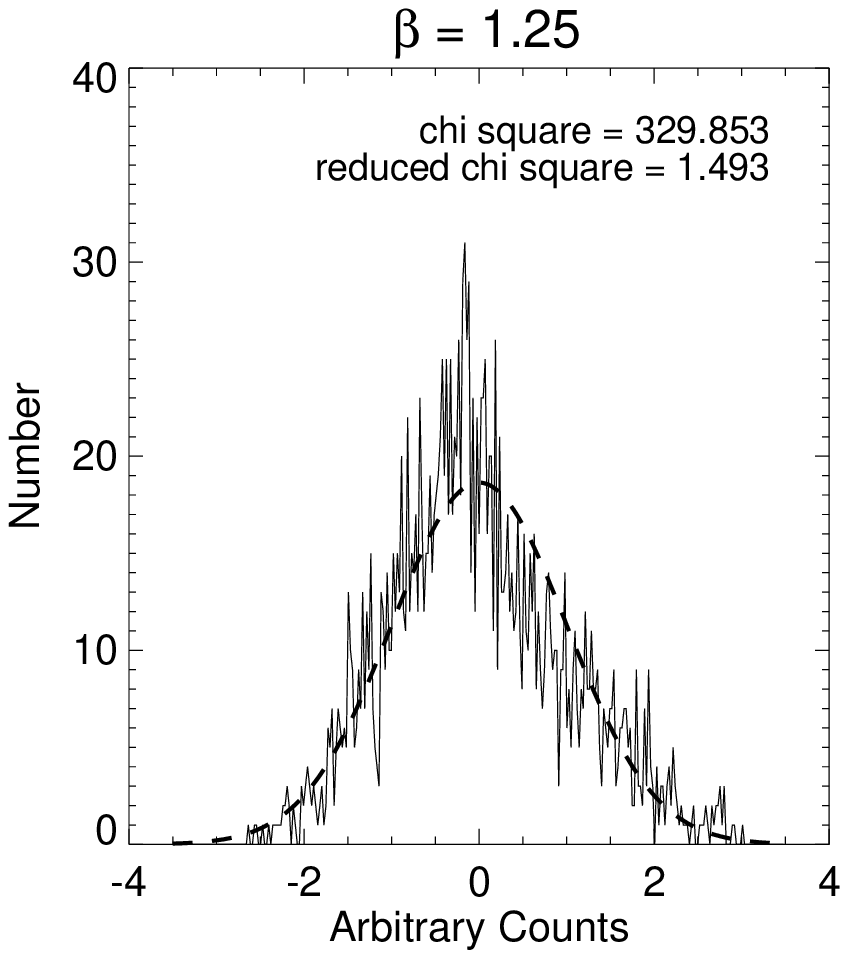}
\epsfxsize = 4.25cm \epsfbox{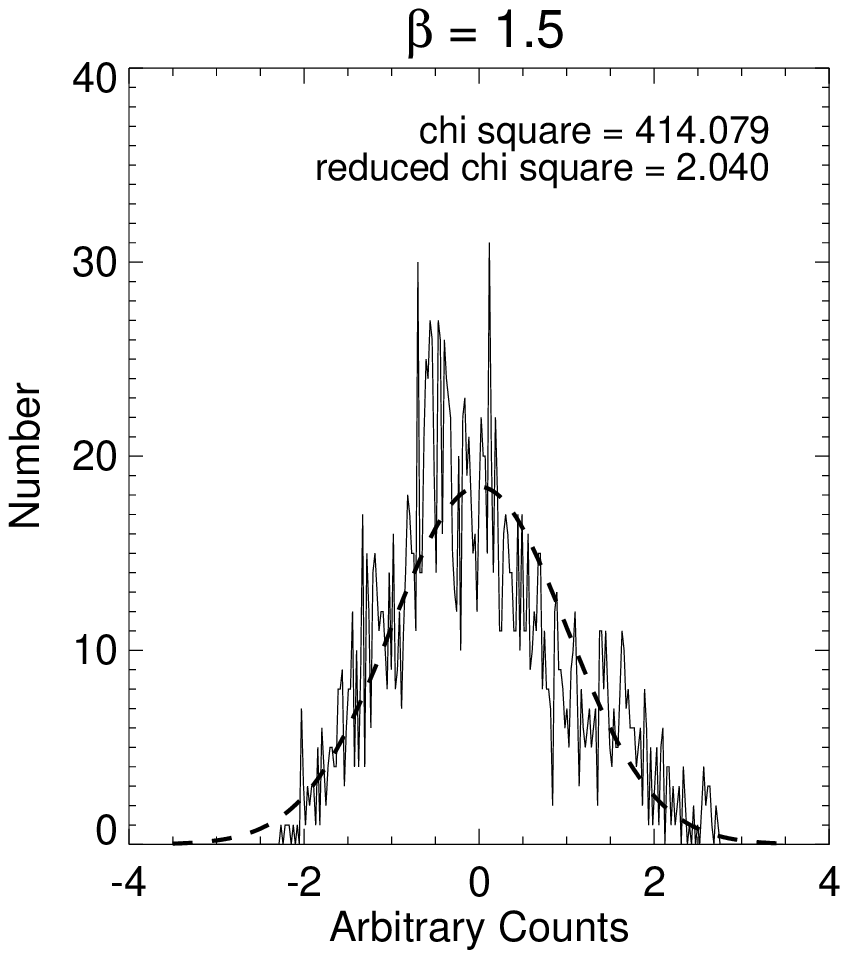}
\epsfxsize = 4.25cm \epsfbox{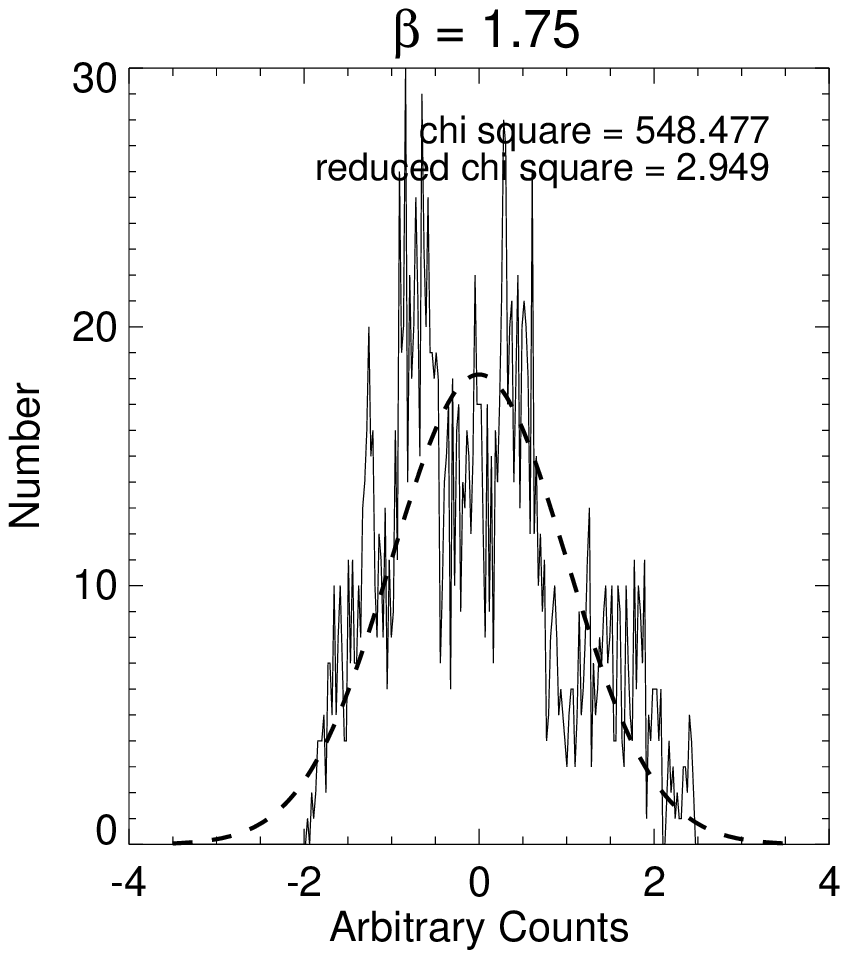}
\epsfxsize = 4.25cm \epsfbox{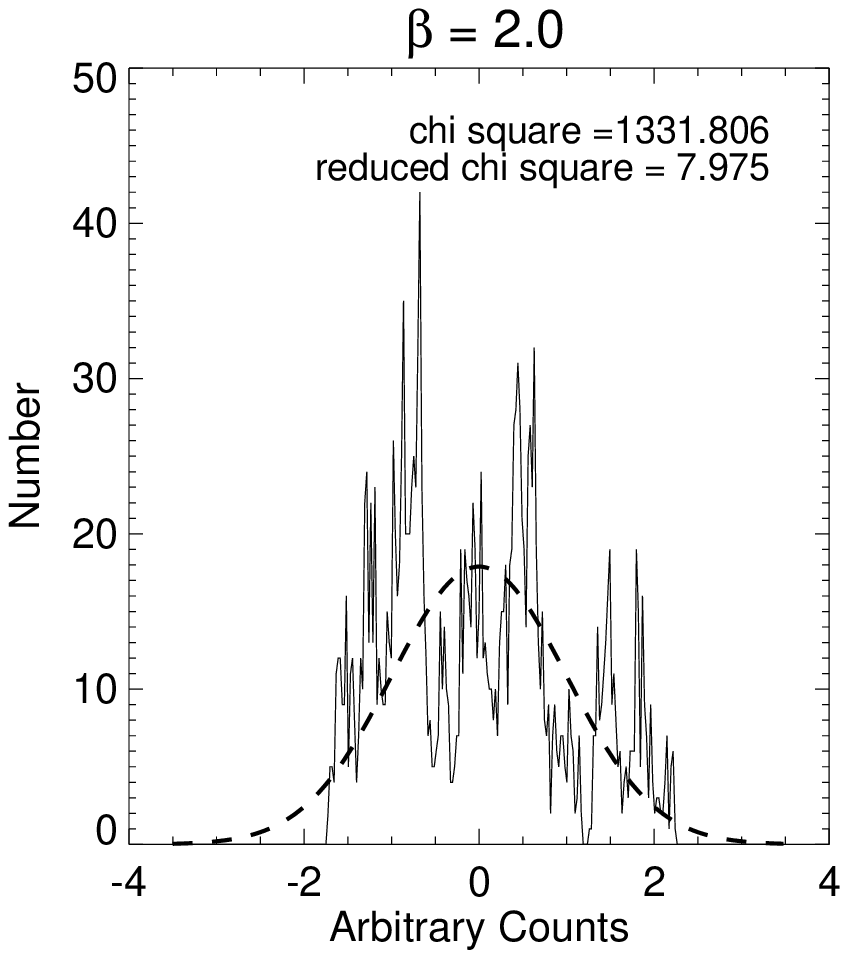} \\
\vspace{3mm}
\caption{Realizations of simulated {\sc flux distributions}, i.e. histograms of simulated lightcurves (continuous lines), together with the best-fitting Gaussian distributions (dashed lines), for all eight non-zero values of $\beta$. Lightcurves, and thus their histograms, are normalized to zero mean and unity standard deviation; accordingly, abscissae are in arbitrary count units. Ordinates denote the number of data points in a given flux bin. \emph{Top group} (above the horizontal line): Results for {\sc run A}; each histogram is composed of 9\,000 flux values. \emph{Bottom group} (below the horizontal line): Results for {\sc run B}; each histogram is composed of 2\,000 flux values. Please note the different axis scales. In each diagram, we quote the ${\chi}^2$ and ${{\chi}_{\rm red}}^2$ values for the Gaussian fits. For $\beta\gtrsim1.25$, the flux distributions break up into several overlapping distributions.}
\label{chi_square}
\end{figure*}

\noindent
For each artificial lightcurve, we obtain its flux distribution by calculating the histogram of the flux values; example results are provided in Fig. \ref{chi_square}. By construction, these flux distributions correspond to the results of single, uniform stochastic emission processes. Depending on the value of $\beta$, we find uni-modal, approximately Gaussian, as well as multi-modal profiles. For $\beta\lesssim1.25$, the flux distributions agree qualitatively with noisy Gaussian distributions. For $\beta\gtrsim1.25$, the distributions become flattened, and multi-modality becomes evident with increasing $\beta$.

In order to check for deviations from uni-modality quantitatively, we probe the agreement of flux histograms with Gaussian distributions by means of a $\chi^2$ test (see also Fig. \ref{chi_square}) with

\begin{equation}
{\chi}^2 = \sum_{i=1}^{N''}\frac{(S_i-E_i)^2}{{Err_i}^2}
\end{equation}

\noindent
(e.g. \citealt{And}) where $S_i$ denotes the $i$th value of the given distribution, $E_i$ denotes the value expected theoretically, and $Err_i$ is the statistical error of the given value. Errors are binomial. For our theoretical Gaussian function we use the parametrization

\begin{equation} \label{model}
f(x;\mu,{\sigma}^2) = A\exp\left[-\frac{1}{2}{\left({\frac{x-\mu}{\sigma}}\right)^2}\right]
\end{equation}

\noindent
where $A$ is the amplitude, $\sigma^2$ is the variance, and $\mu$ is the mean of the distribution. Because of our normalization of the simulated flux distributions (Eq. \ref{norm}), we have $\sigma^2=1$ and $\mu=0$, leaving $A$ as the only free parameter. We also obtain the reduced ${\chi}^2$ given by

\begin{equation}
\chi^2_{\rm red} = \frac{{\chi}^2}{DOF}
\end{equation}

\noindent
where $DOF$ is the number of degrees of freedom of the model given by

\begin{equation}
DOF = N'' - P
\end{equation}

\noindent
(e.g. \citealt{And}) where $N''$ is the number of bins of the histogram that have non-zero values and $P$ is the number of parameters of the model. In our case, $P=1$ because the amplitude of the Gaussian function is our only model parameter (see Eq. \ref{model}). If the observed distribution corresponds to the expected one within errors, $\chi^2_{\rm red}\approx1$. We apply $\chi^2$ tests to each simulated flux distribution for each value of $\beta$; example results of these tests are shown in Fig. \ref{chi_square}. We observe a systematic increase of $\chi_{\rm red}^2$ with increasing $\beta$, suggesting substantial deviation from Gaussian distributions occurring for $\beta>1$.

\begin{figure*}
\centering
\epsfxsize = 4.25cm \epsfbox{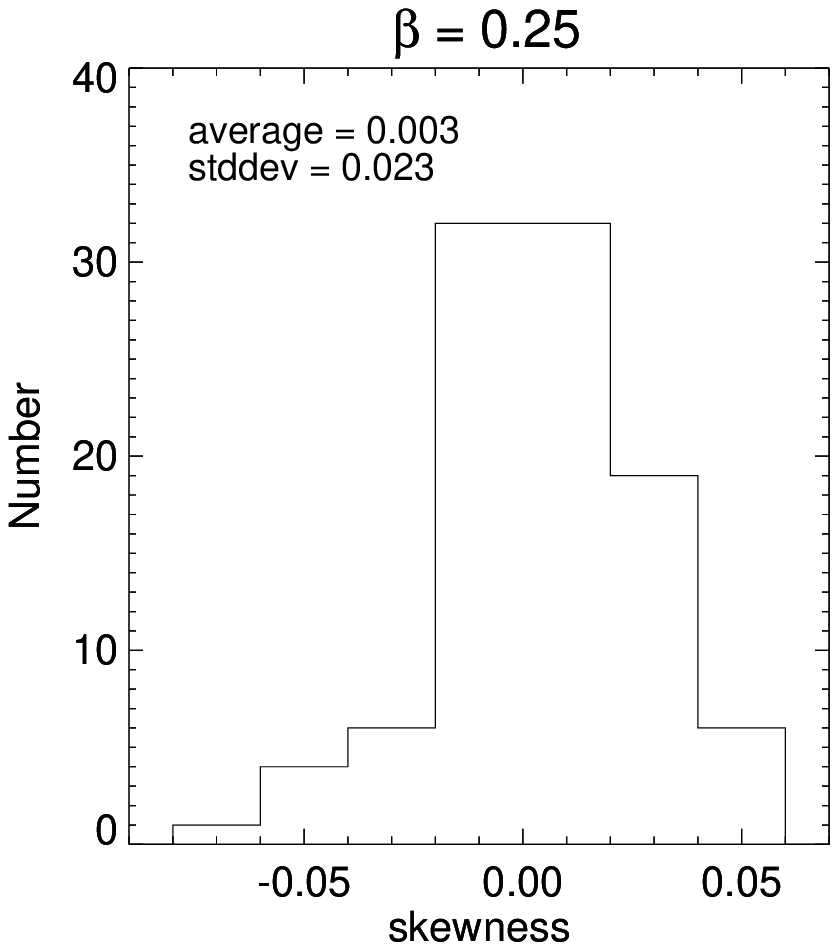}
\epsfxsize = 4.25cm \epsfbox{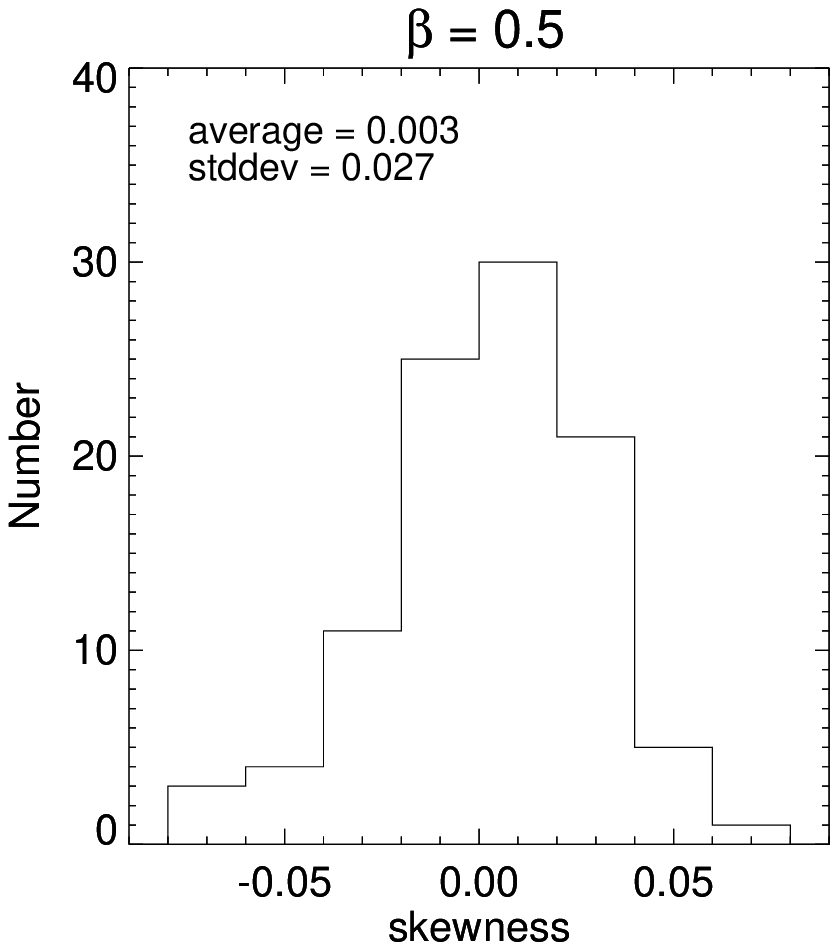}
\epsfxsize = 4.25cm \epsfbox{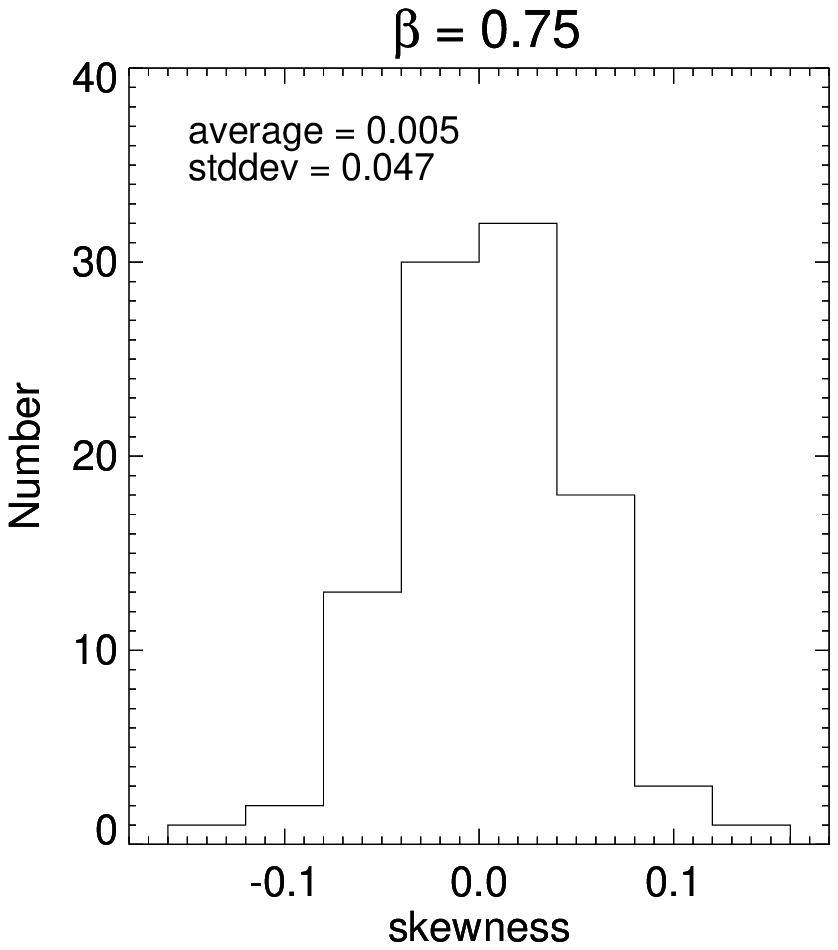}
\epsfxsize = 4.25cm \epsfbox{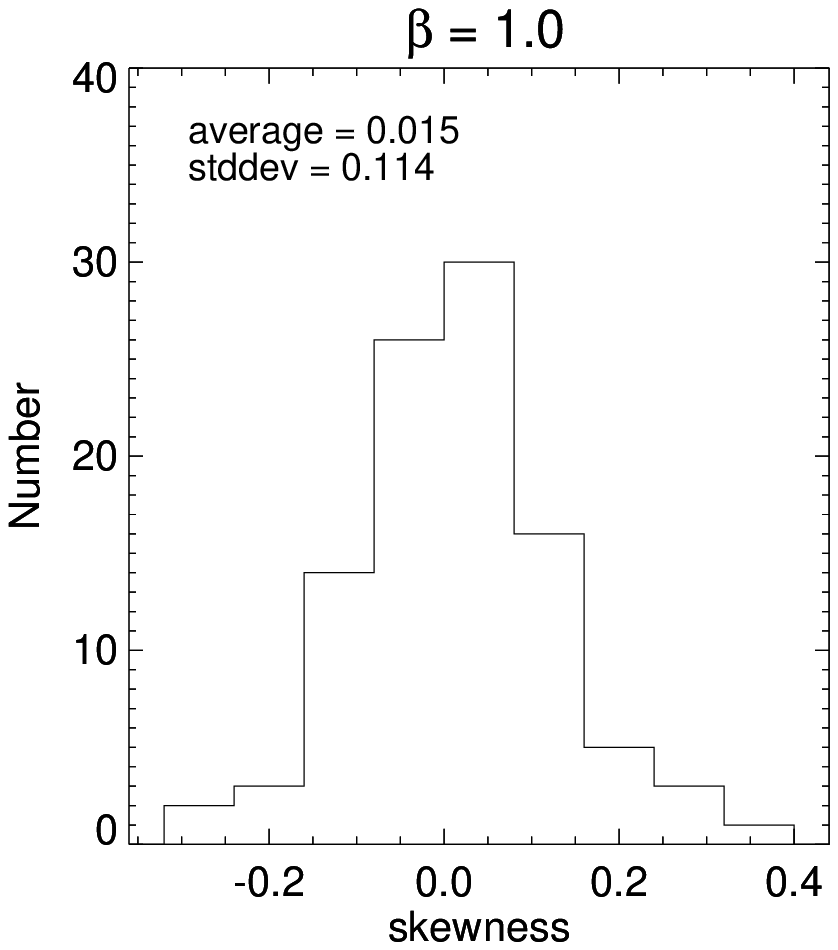} \\
\vspace{2mm}
\epsfxsize = 4.25cm \epsfbox{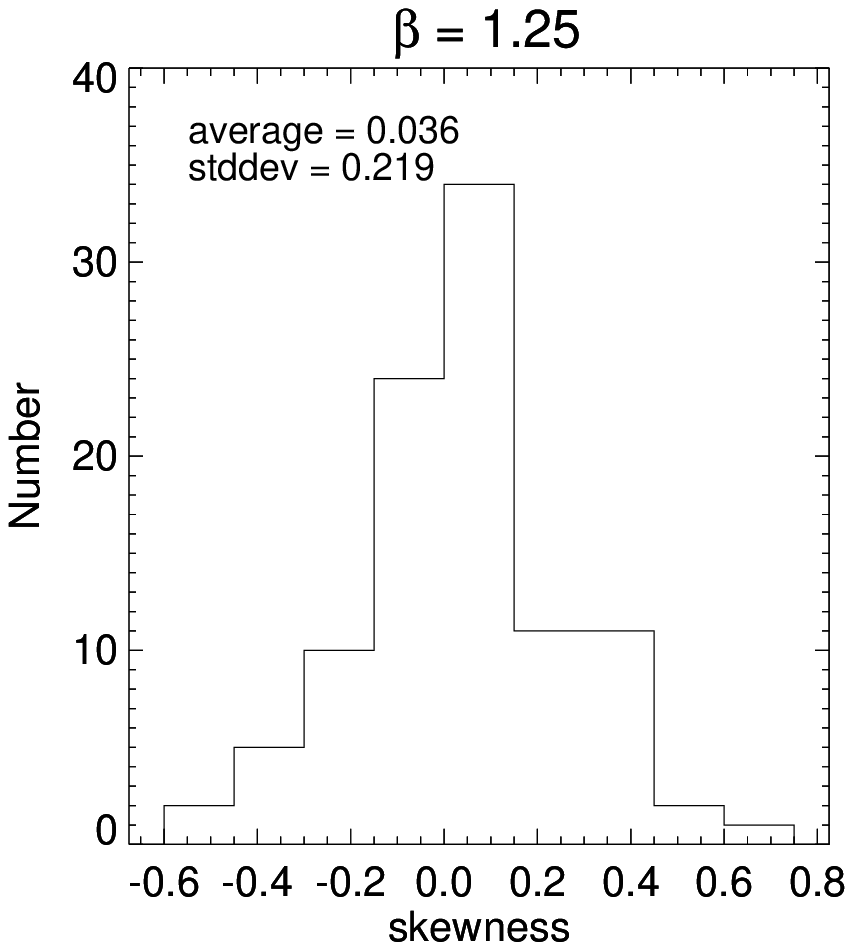}
\epsfxsize = 4.25cm \epsfbox{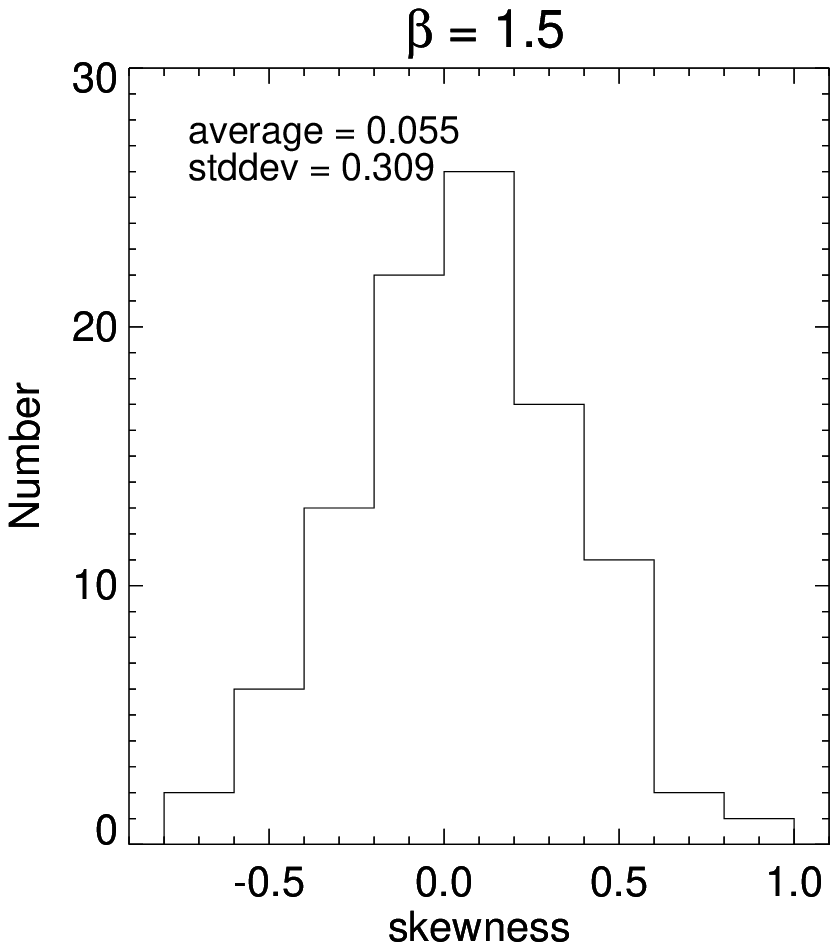}
\epsfxsize = 4.25cm \epsfbox{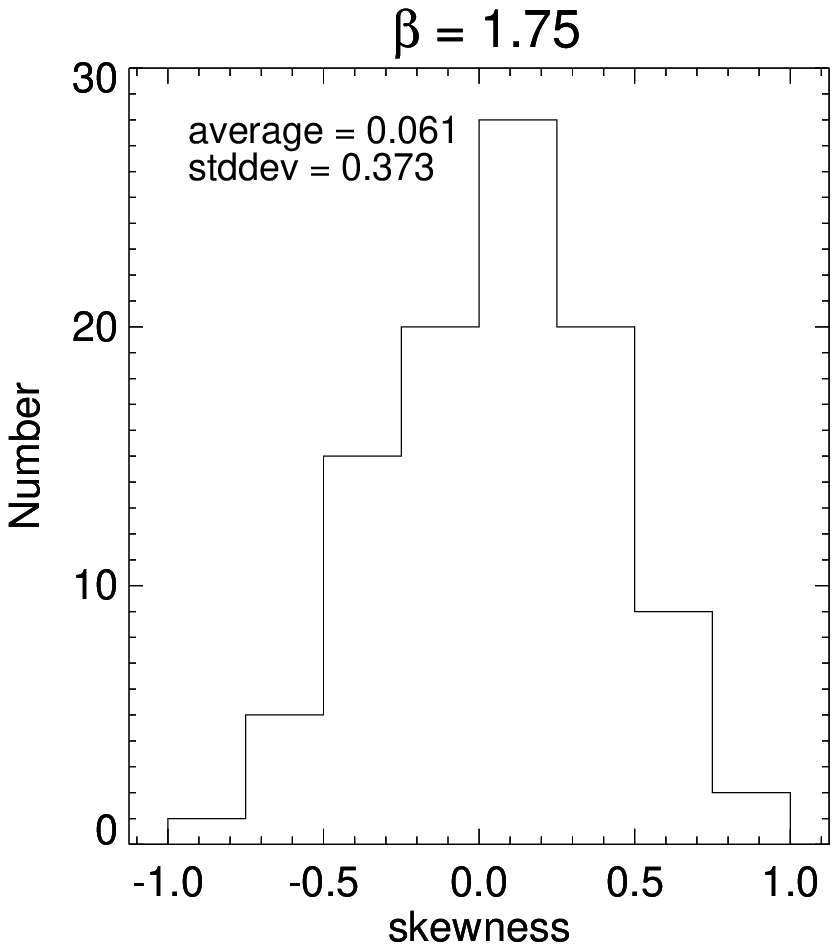}
\epsfxsize = 4.25cm \epsfbox{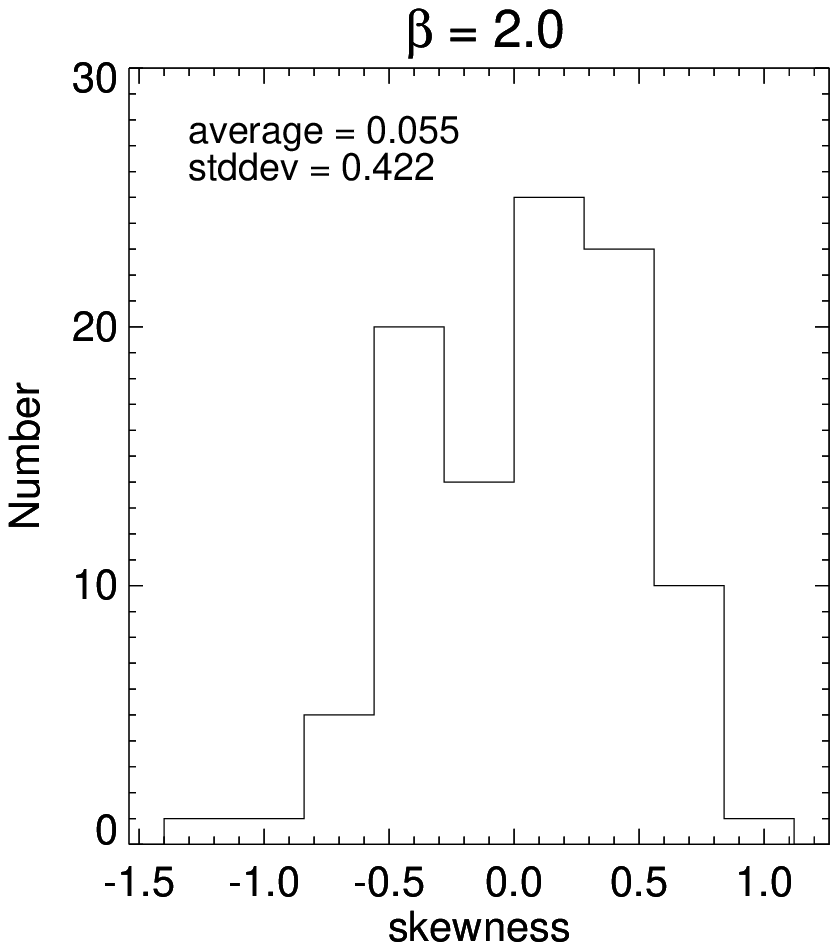} \\
\hrulefill \\
\vspace{6mm}
\epsfxsize = 4.25cm \epsfbox{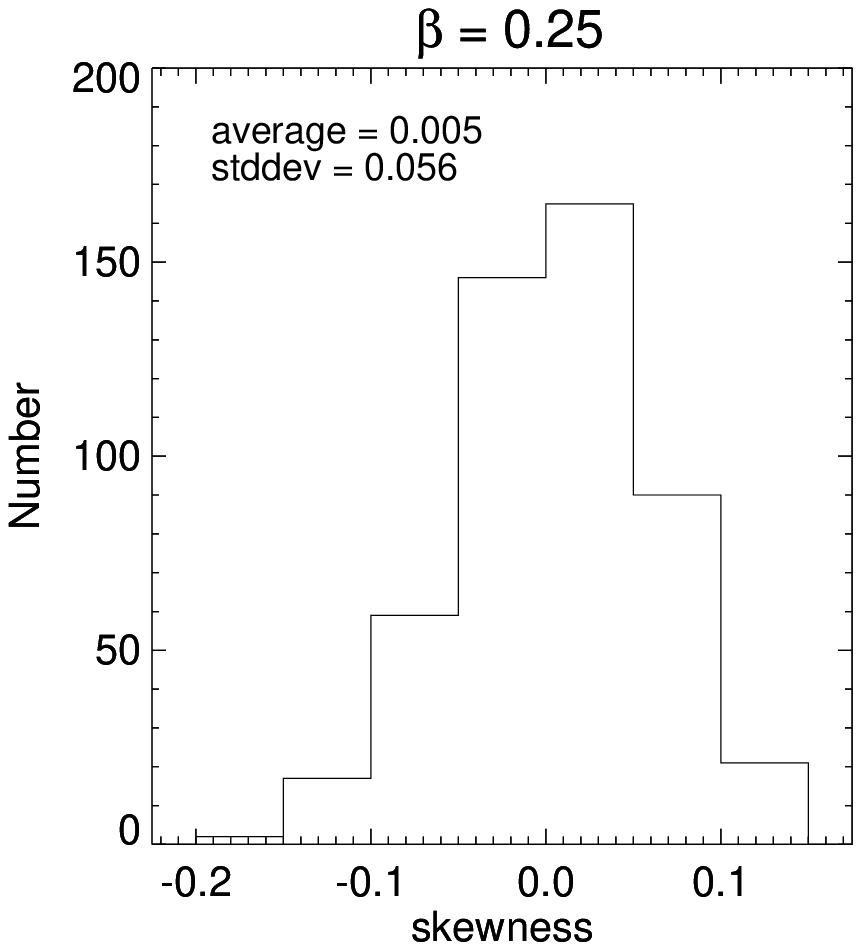}
\epsfxsize = 4.25cm \epsfbox{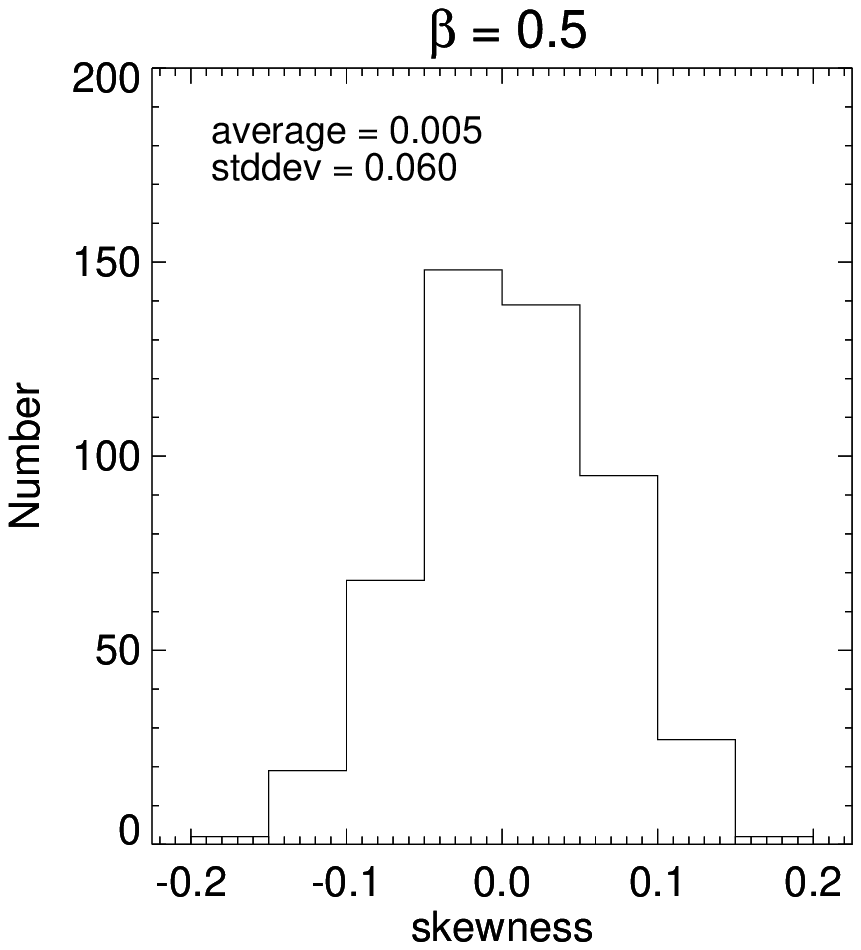}
\epsfxsize = 4.25cm \epsfbox{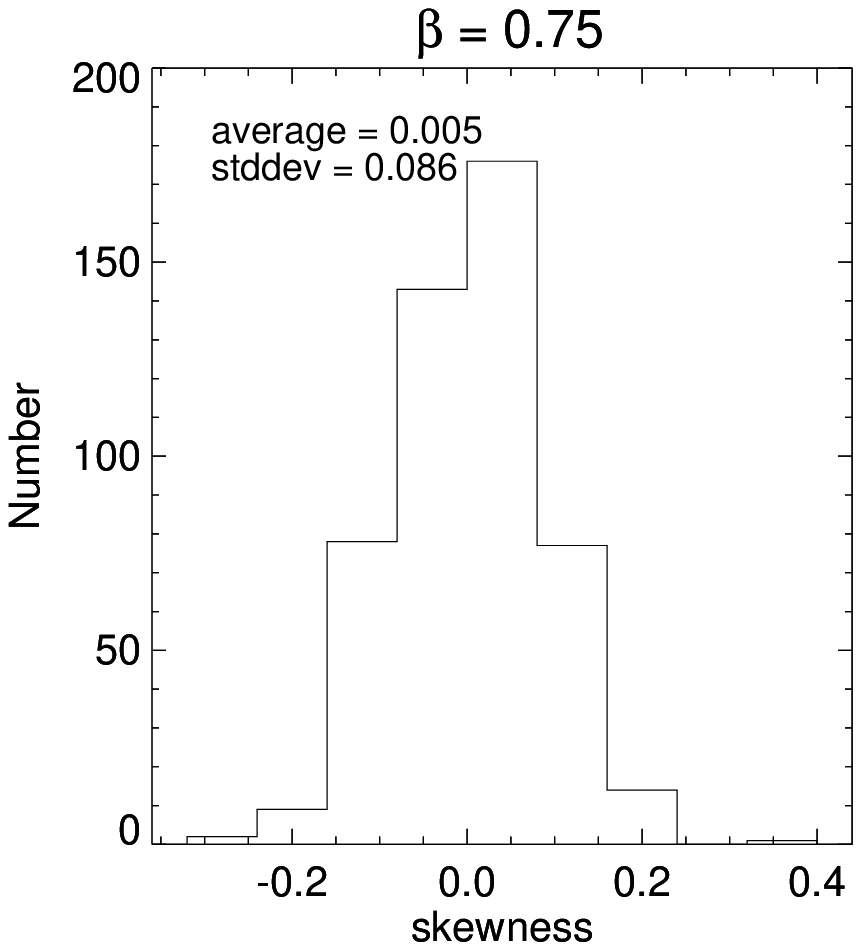}
\epsfxsize = 4.25cm \epsfbox{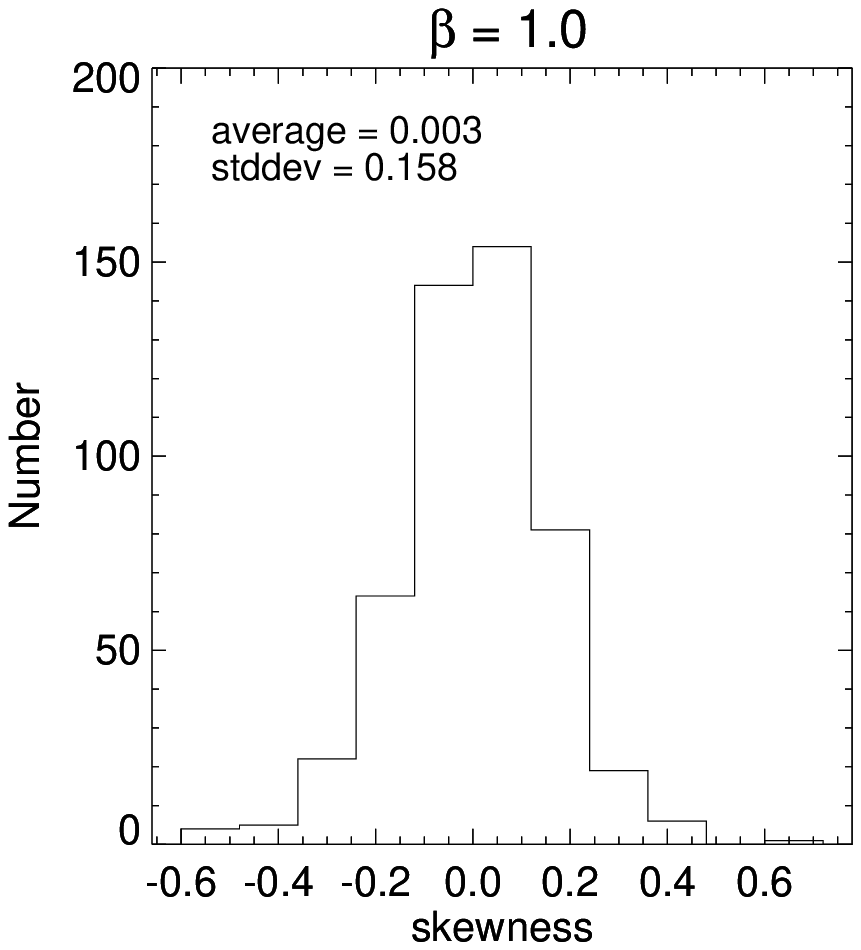} \\
\vspace{2mm}
\epsfxsize = 4.25cm \epsfbox{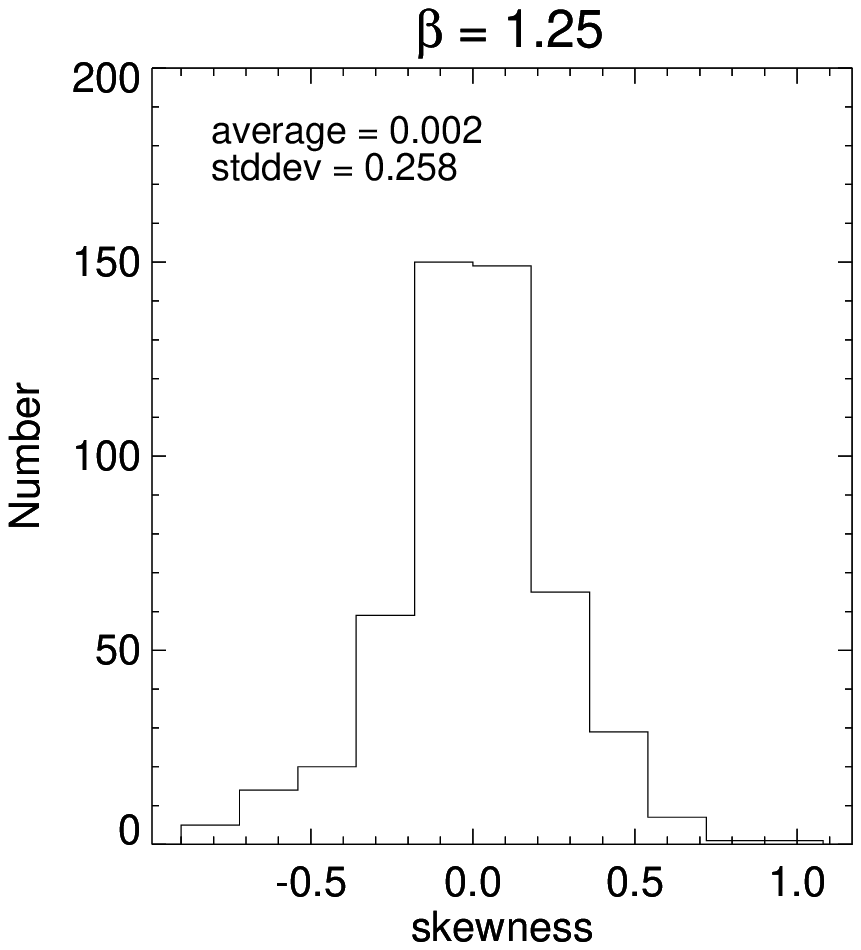}
\epsfxsize = 4.25cm \epsfbox{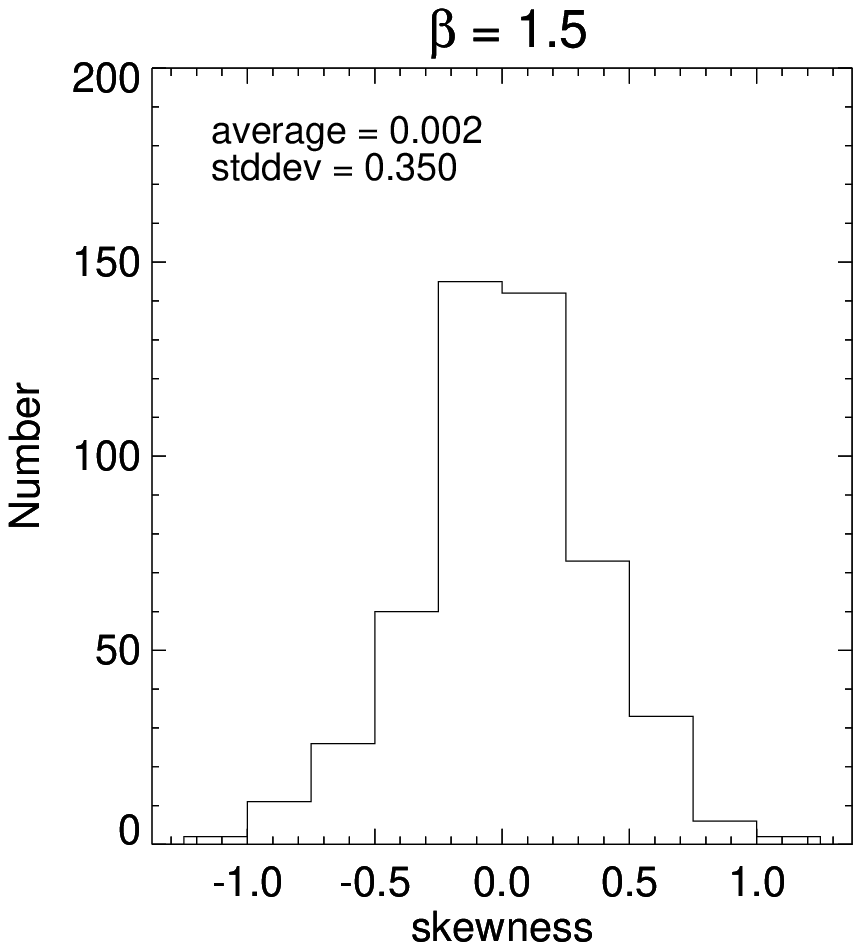}
\epsfxsize = 4.25cm \epsfbox{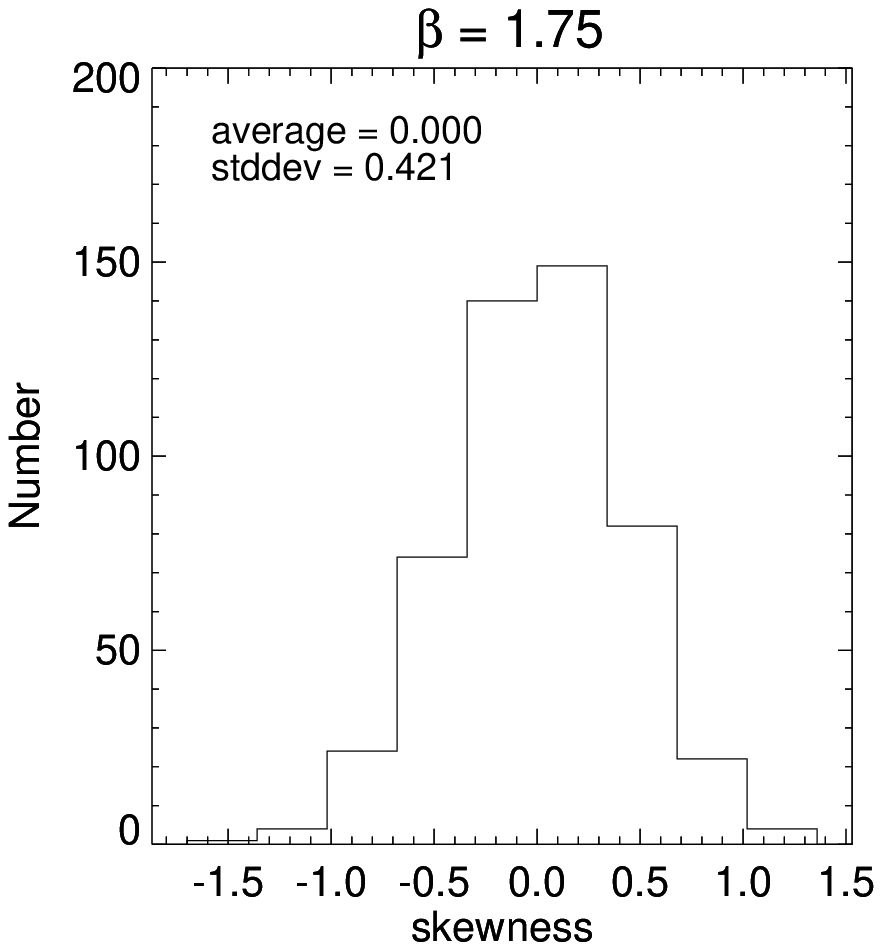}
\epsfxsize = 4.25cm \epsfbox{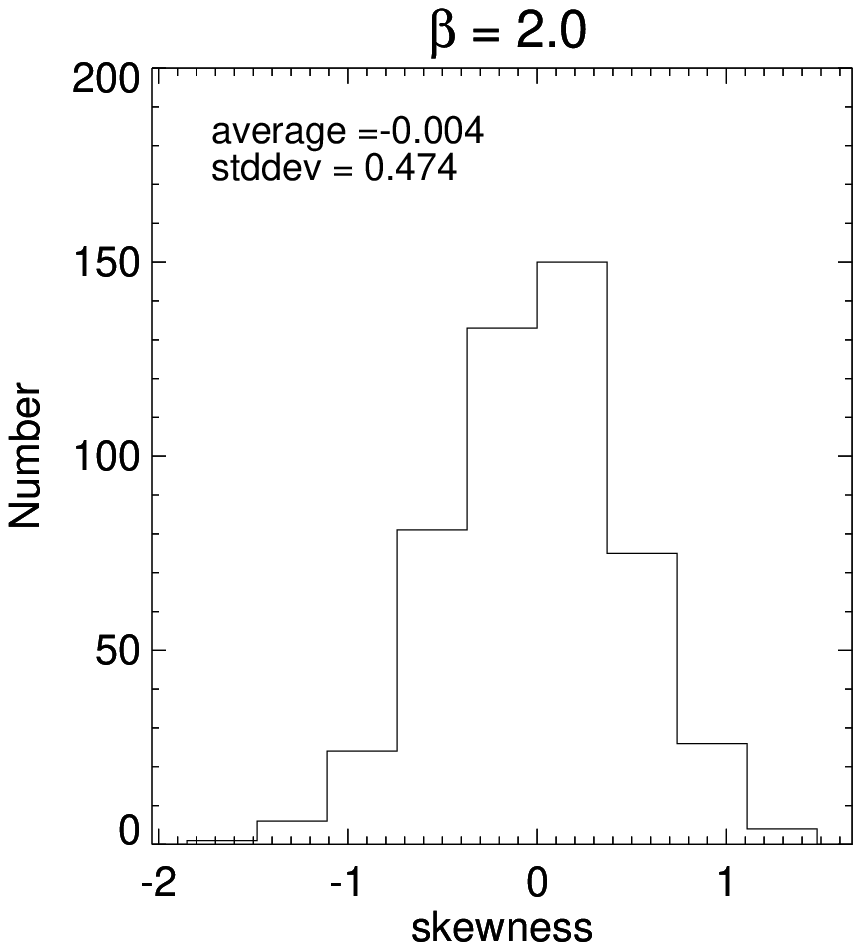} \\
\vspace{3mm}
\caption{Histograms of the {\sc skewness} of the simulated flux distributions for all eight non-zero values of $\beta$. (Examples for flux distributions are shown in Fig.~\ref{chi_square}.) Abscissae indicate the (binned) skewness values observed; ordinates show the number of simulated flux distributions with a skewness located in a given bin. \emph{Top group} (above the horizontal line): Results for {\sc run A}; each histogram includes the results from 100 trials. \emph{Bottom group} (below the horizontal line): Results for {\sc run B}; each histogram includes the results from 500 trials. Please note the different axis scales. The average values of skewness do not significantly -- in terms of standard error of mean -- deviate from zero, meaning that the given functions agree with Gaussian profiles with respect to symmetry.}
\label{skewness}
\end{figure*}

\begin{figure*}
\centering
\epsfxsize = 4.25cm \epsfbox{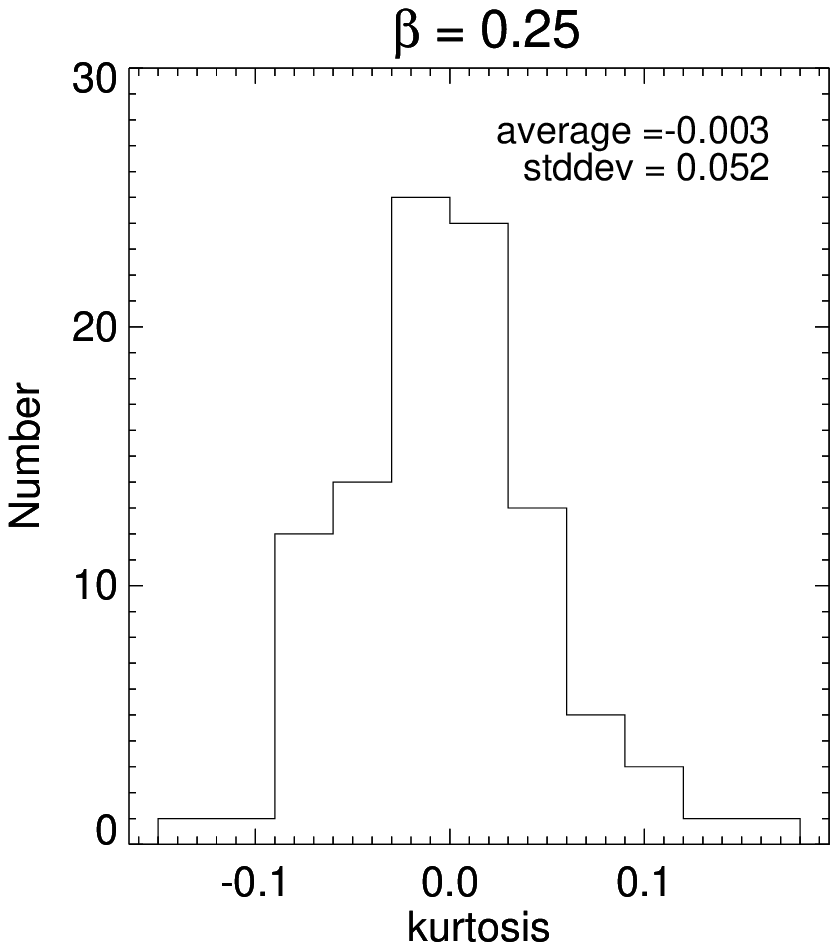}
\epsfxsize = 4.25cm \epsfbox{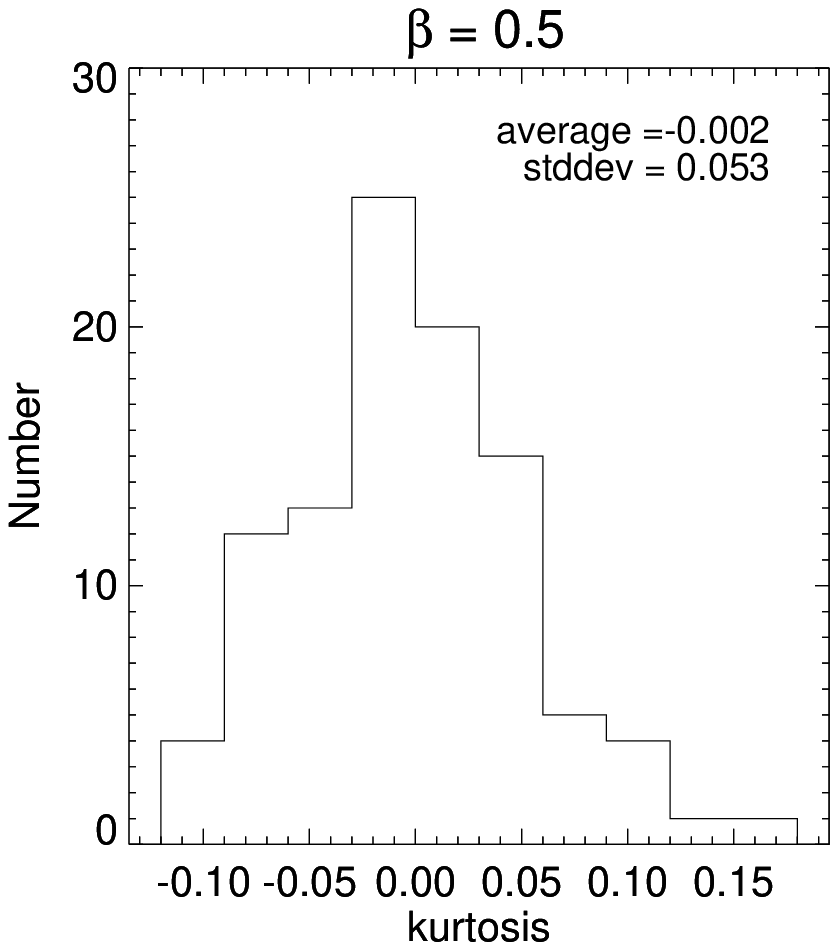} 
\epsfxsize = 4.25cm \epsfbox{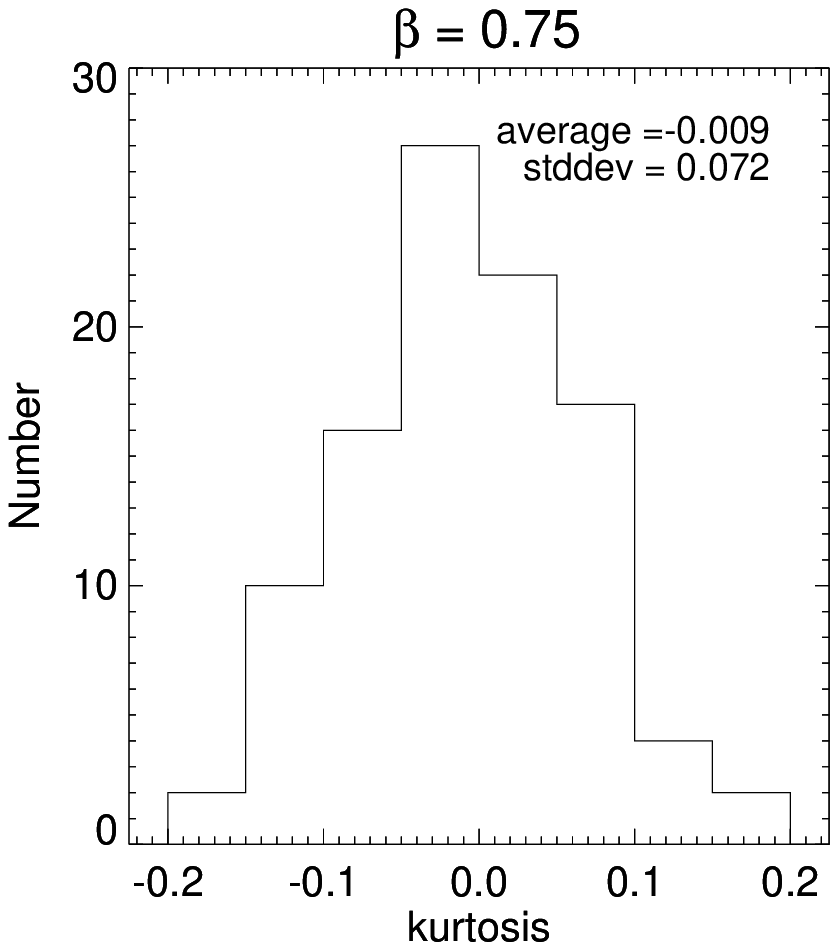}
\epsfxsize = 4.25cm \epsfbox{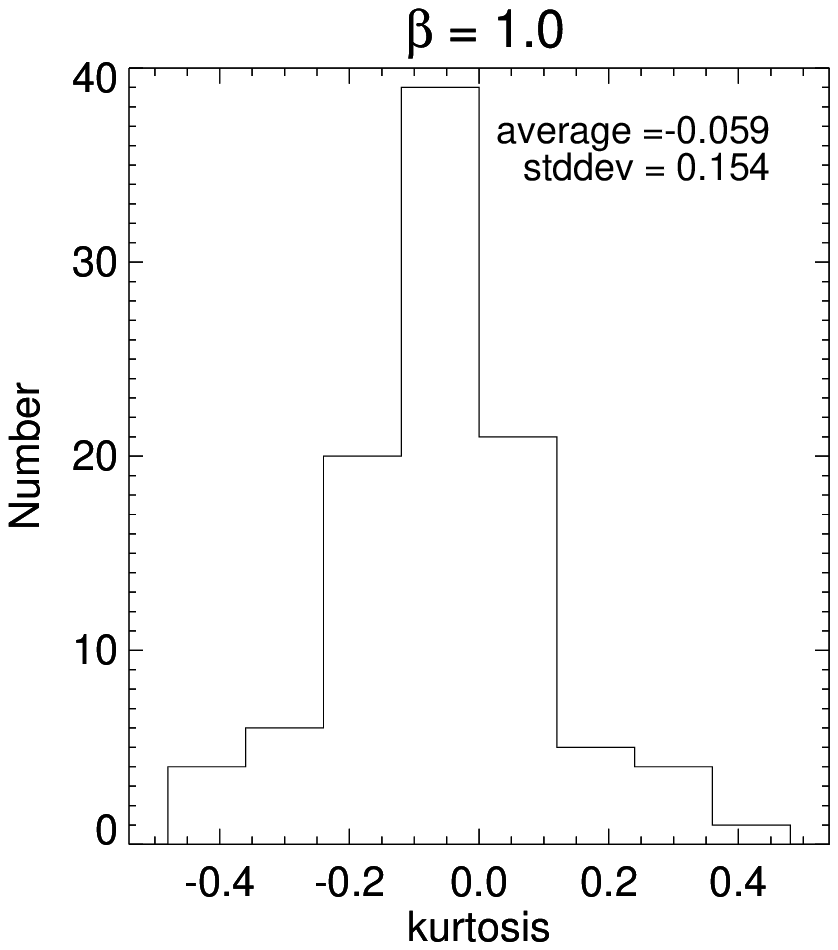} \\
\vspace{2mm}
\epsfxsize = 4.25cm \epsfbox{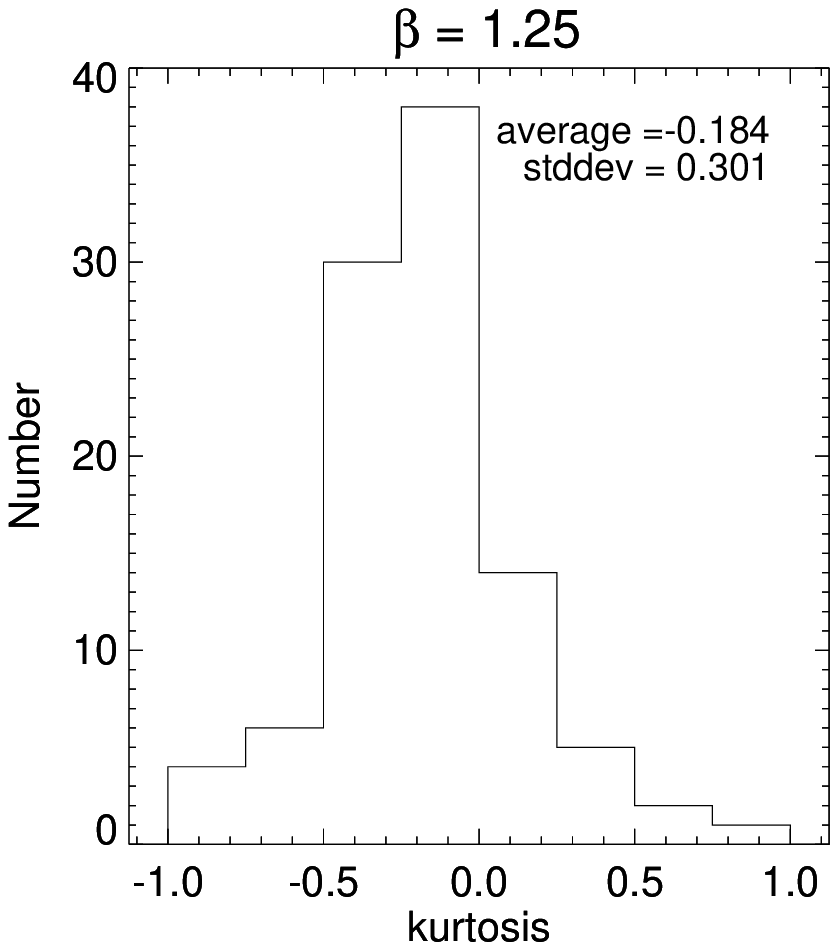} 
\epsfxsize = 4.25cm \epsfbox{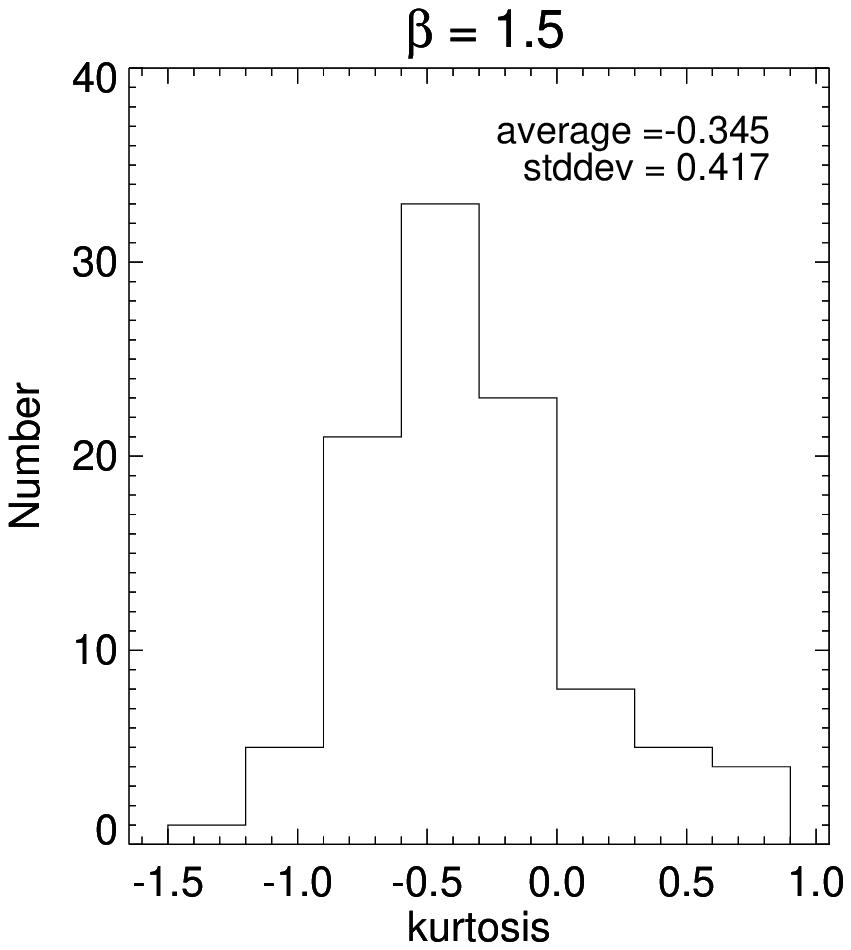}
\epsfxsize = 4.25cm \epsfbox{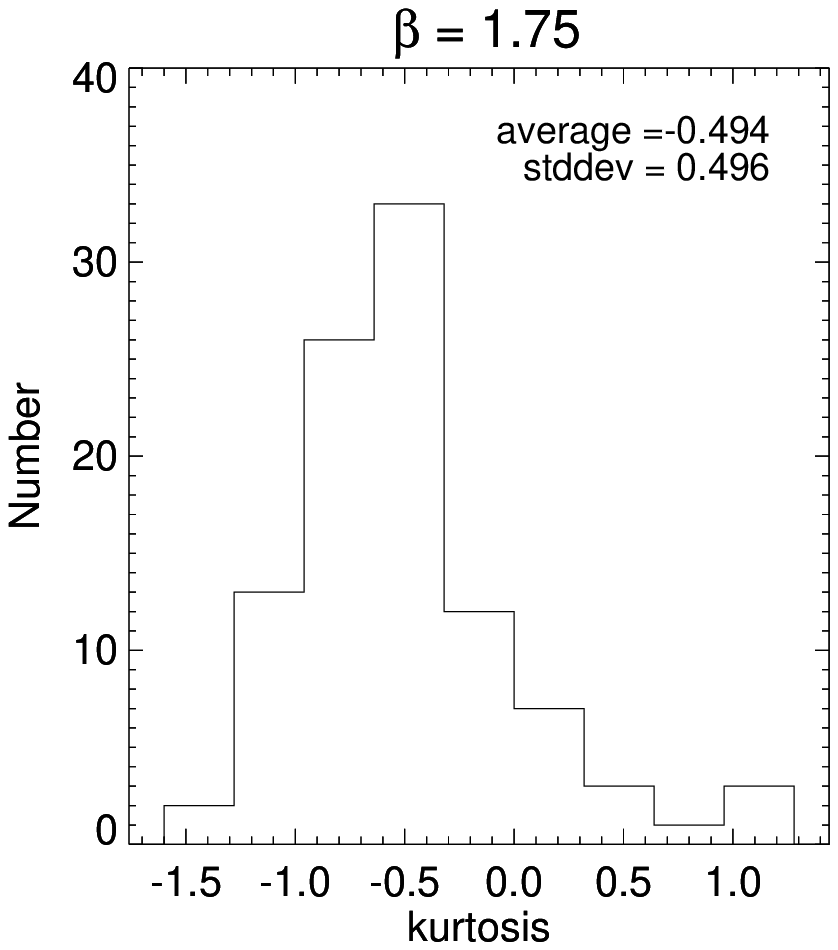}
\epsfxsize = 4.25cm \epsfbox{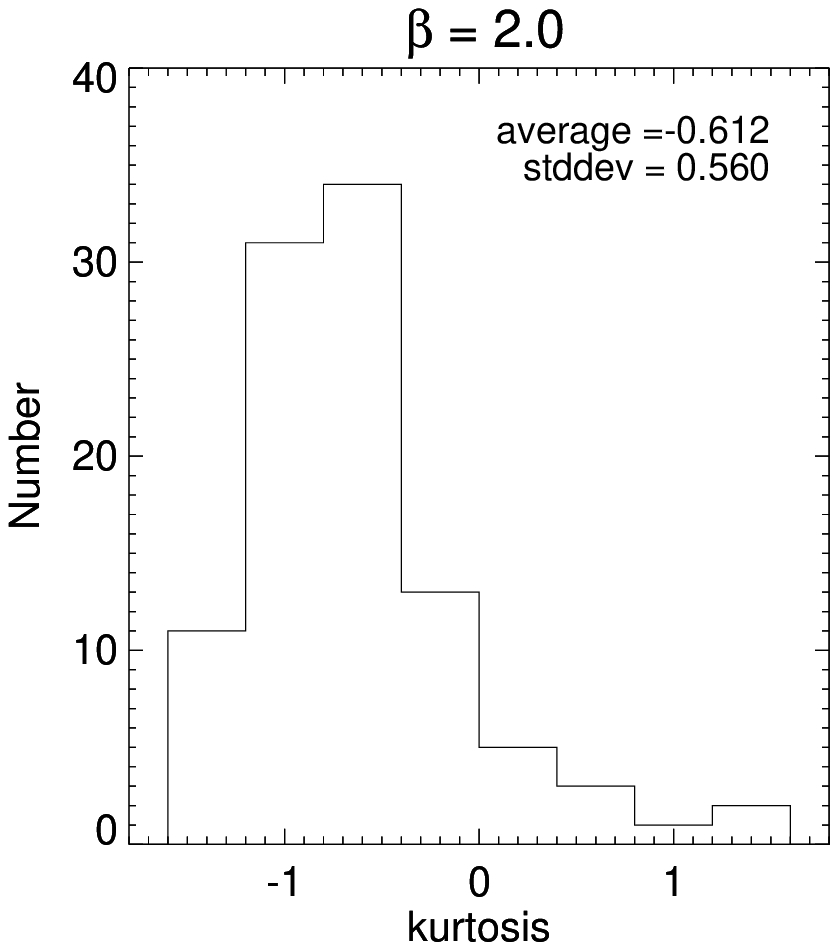} \\
\hrulefill \\
\vspace{6mm}
\epsfxsize = 4.25cm \epsfbox{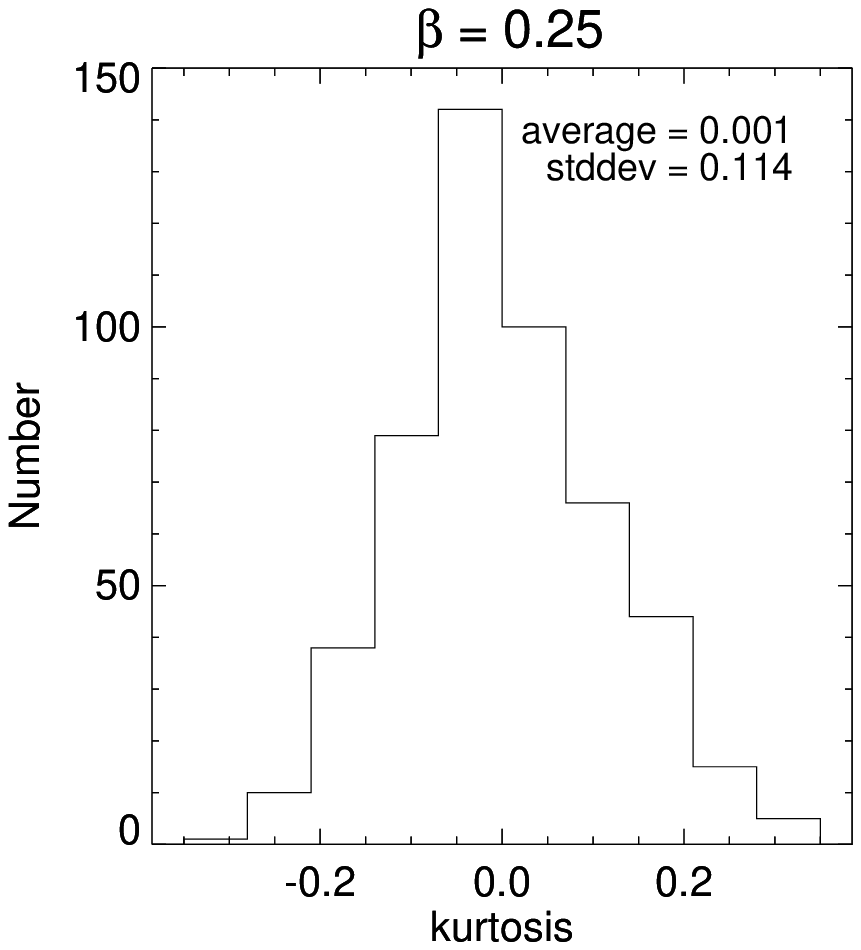}
\epsfxsize = 4.25cm \epsfbox{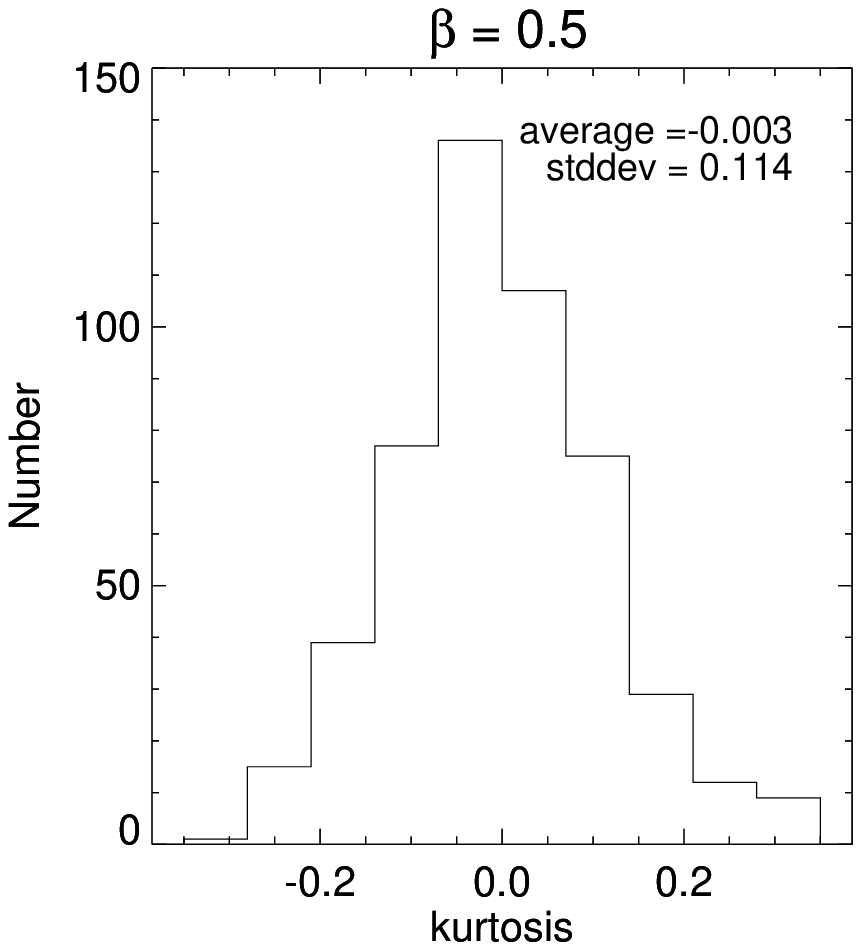} 
\epsfxsize = 4.25cm \epsfbox{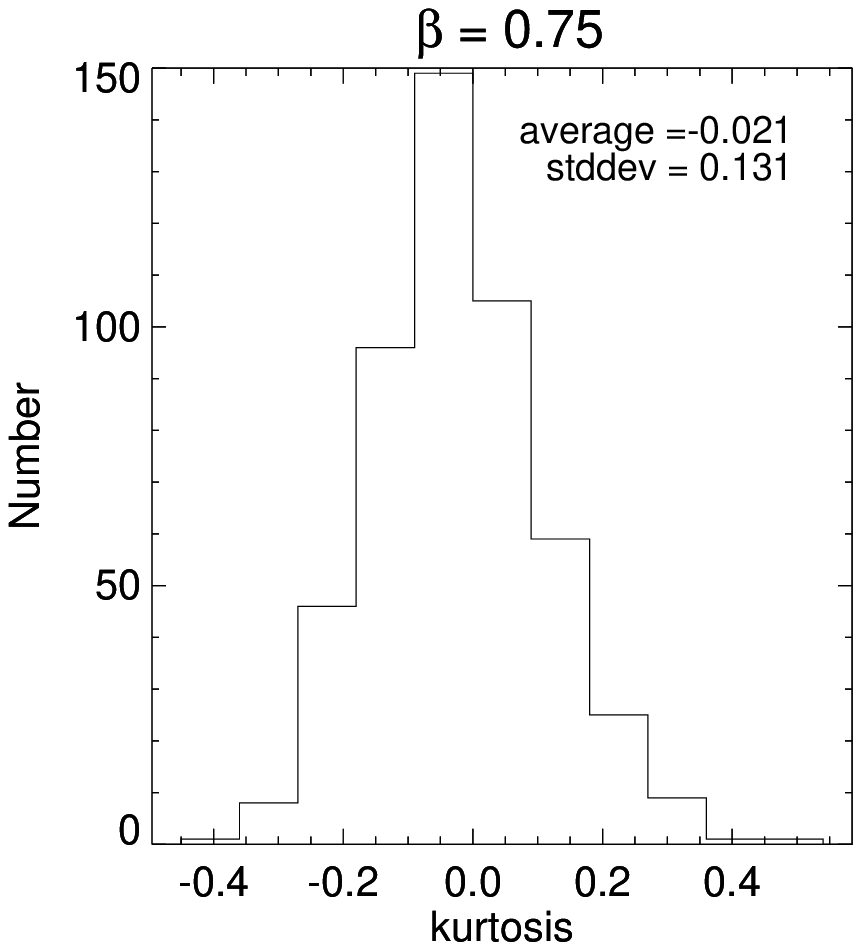}
\epsfxsize = 4.25cm \epsfbox{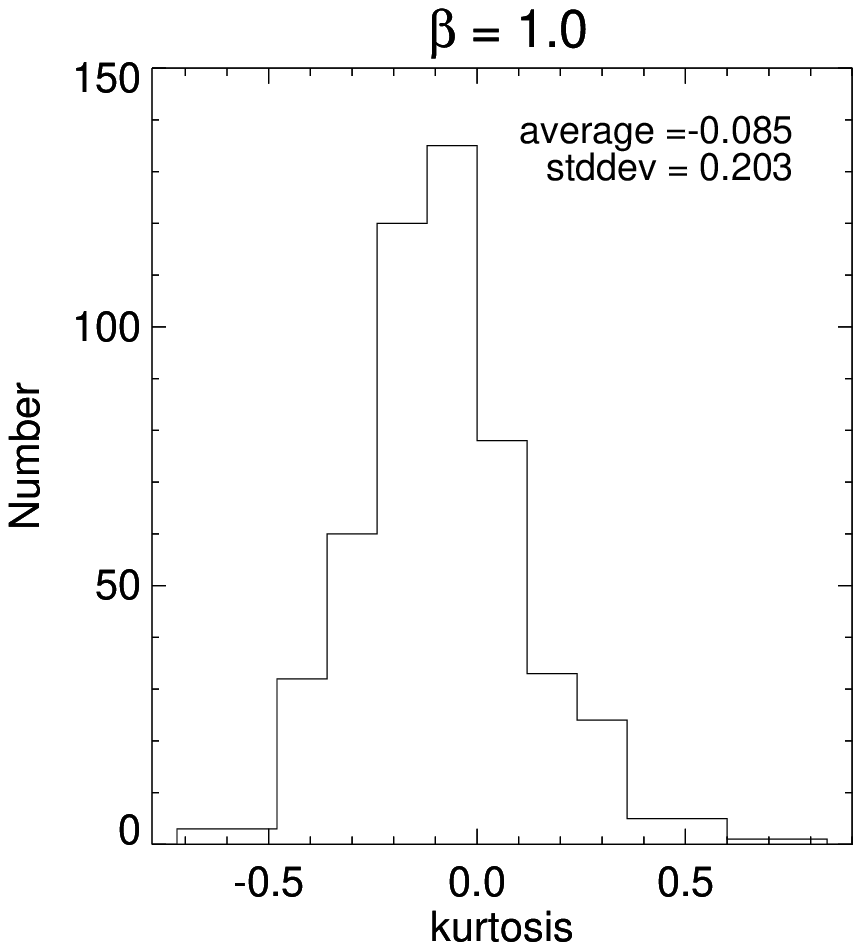} \\
\vspace{2mm}
\epsfxsize = 4.25cm \epsfbox{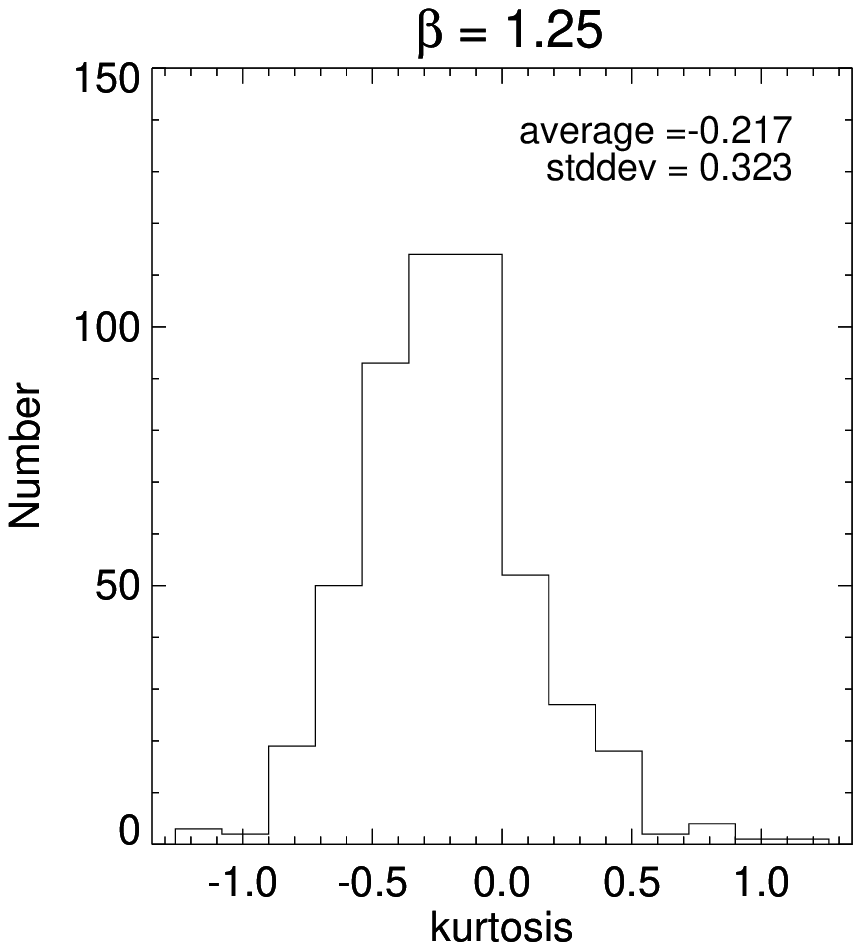} 
\epsfxsize = 4.25cm \epsfbox{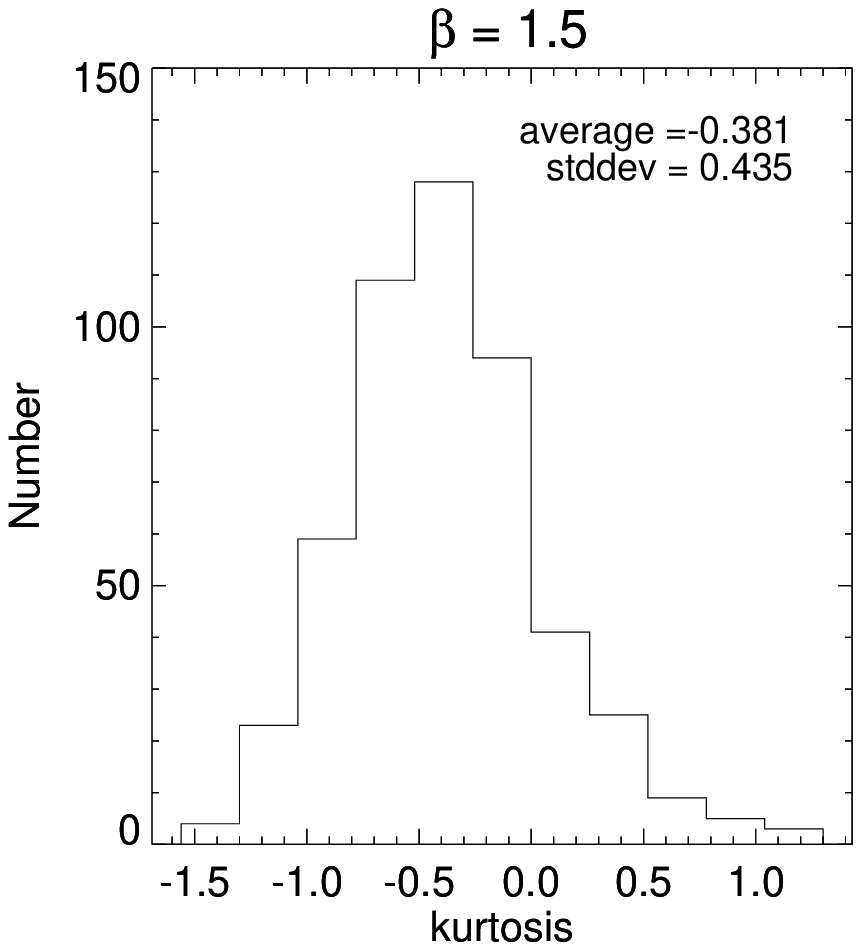}
\epsfxsize = 4.25cm \epsfbox{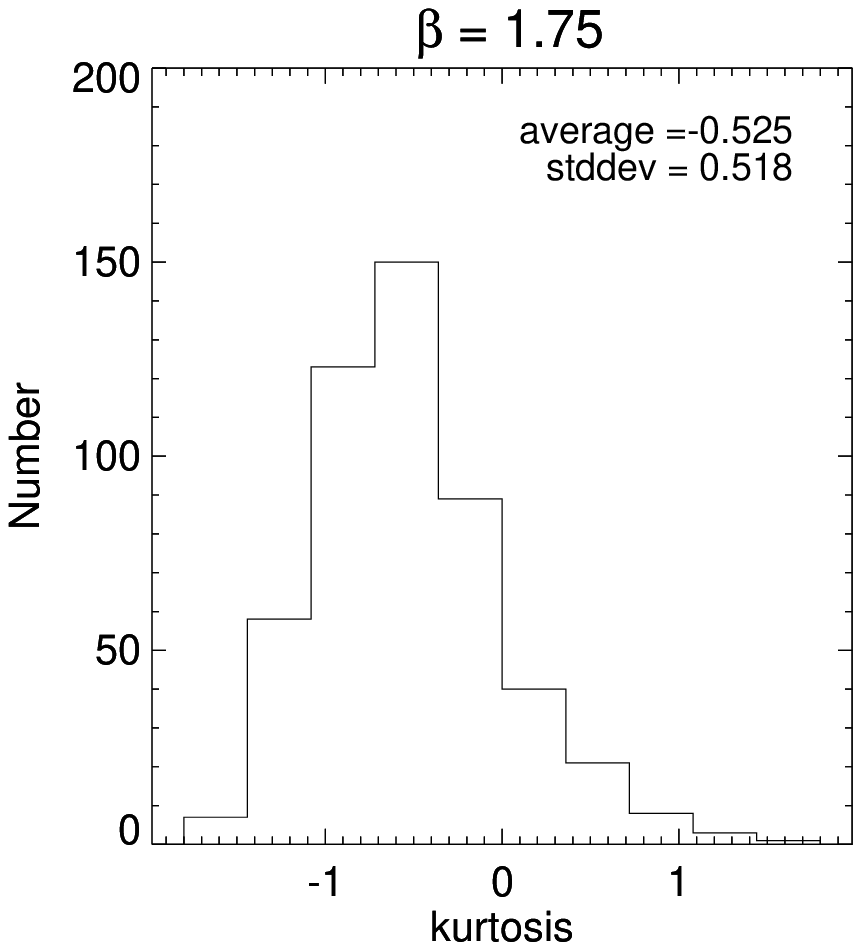}
\epsfxsize = 4.25cm \epsfbox{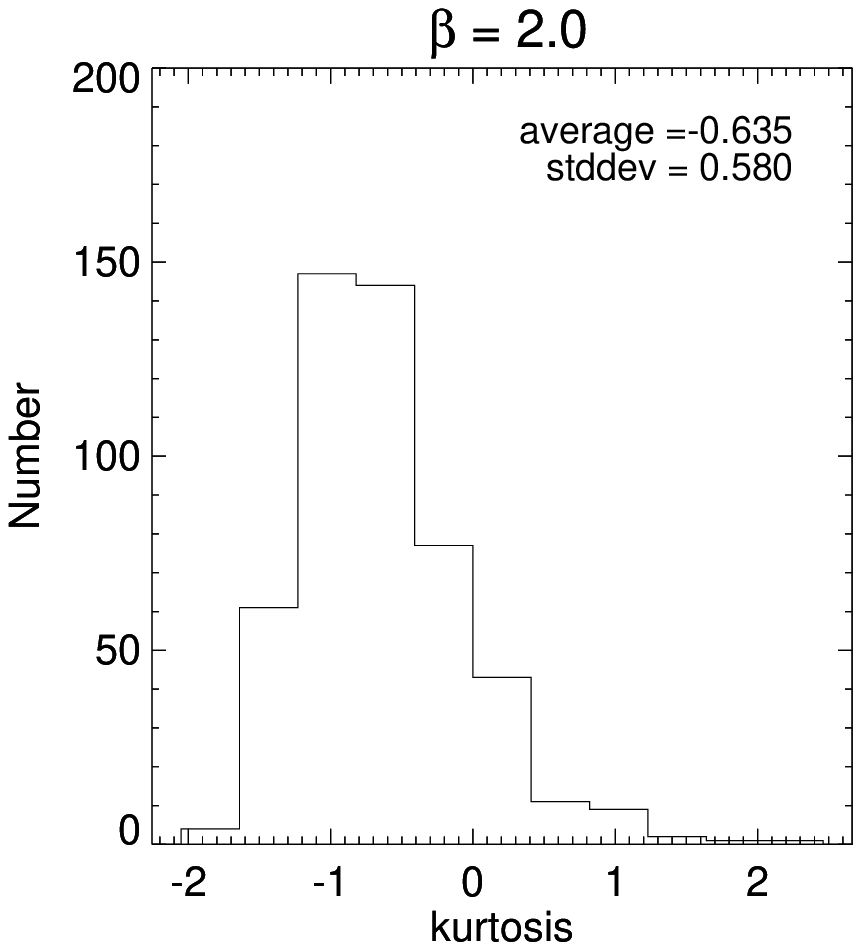} \\
\vspace{3mm}
\caption{Histograms of the {\sc kurtosis} of the simulated flux distributions for all eight non-zero values of $\beta$. Abscissae indicate the (binned) kurtosis values observed; ordinates show the number of simulated flux distributions with a kurtosis located in a given bin. \emph{Top group} (above the horizontal line): Results for {\sc run A}; each histogram includes the results from 100 trials. \emph{Bottom group} (below the horizontal line): Results for {\sc run B}; each histogram includes the results from 500 trials. Please note the different axis scales. The average values are significantly -- in terms of standard error of mean -- smaller than zero for $\beta\gtrsim1.0$, meaning that the observed distributions are too flat compared to Gaussian profiles (platykurtic distributions). This signature is expected for multi-modal flux distributions.}
\label{kurtosis}
\end{figure*}

\begin{deluxetable}{lccccccccc}
\tablecolumns{10}
\tablewidth{0pc}
\tablecaption{{\sc Skewness} statistics for {\sc runs A and B}. Deviations from zero are in units of $\sigma_{\overline x}$, i.e. the standard error of mean. \label{tbl1}}
\tablehead{
\colhead{$\beta$} & 
\colhead{$0.0$} & \colhead{$0.25$} & \colhead{$0.5$} & \colhead{$0.75$} & \colhead{$1.0$} 
& \colhead{$1.25$} & \colhead{$1.5$} & \colhead{$1.75$} & \colhead{$2.0$}
}
\startdata
{\sc Run A:} & & & & & & & & & \\
mean & 0.003 & 0.003 & 0.003 & 0.005 & 0.015 & 0.036 & 0.055 & 0.061 & 0.055\\
standard deviation & 0.023 & 0.023 & 0.027 & 0.047 & 0.114 & 0.219 & 0.309 & 0.373 & 0.422\\
standard error & 0.002 & 0.002 & 0.003 & 0.005 & 0.011 & 0.022 & 0.031 & 0.037 & 0.042\\
deviation from 0 ($\sigma_{\overline x}$) & 1.261 & 1.335 & 1.250 & 1.085 & 1.272 & 1.635 & 1.774 & 1.628 & 1.302\\
&&&&&&&&& \\
{\sc Run B:} & & & & & & & & & \\
mean & 0.004 & 0.005 & 0.005 & 0.005 & 0.003 & 0.002 & 0.002 & 0.000 & $-$0.004 \\
standard deviation & 0.056 & 0.056 & 0.060 & 0.086 & 0.158 & 0.258 & 0.350 & 0.421 & 0.474\\
standard error & 0.003 & 0.003 & 0.003 & 0.004 & 0.007 & 0.012 & 0.016 & 0.019 & 0.021\\
deviation from 0 ($\sigma_{\overline x}$) & 1.597 & 1.996 & 1.863 & 1.300 & 0.425 & 0.173 & 0.128 & 0.000 & 0.189\\
\enddata
\end{deluxetable}

Whereas a $\chi^2$ test provides information on the presence of deviations between model and data, it does not tell us \emph{how} model and data deviate. We are here specifically concerned about a ``breakup'' of a single flux distribution into multiple, potentially overlapping, distributions with increasing values of $\beta$ as illustrated in Fig. \ref{chi_square}. Accordingly, we find distributions that can deviate systematically from Gaussians in terms of symmetry as well as peakedness (concentration).

The degree of asymmetry of a probability distribution is quantified by the \emph{skewness}, or third standardized moment, 

\begin{equation}
g_1=\frac{\mu_3}{\sigma^3}=\frac{E\left[(X-\mu)^3\right]}{(E[(X-\mu)^2])^{3/2}}
\label{eq_skew}
\end{equation}

\noindent
(e.g. \citealt{craw,press2007}). Here $E$ is the expectation value operator, $X$ is the data set, $\mu_3$ is the third moment about the mean $\mu$, and $\sigma$ is the standard deviation. If the probability distribution is Gaussian, the skewness is zero by construction. For each value of $\beta$, we calculate $g_1$ for all $N'$ realizations of flux distributions; we present the corresponding histograms in Fig. \ref{skewness}. Within errors, the average values of skewness agree with zero for all $\beta$. We note, however, that the width of the skewness distributions -- quantified by the standard deviations of the histograms -- increases with increasing $\beta$, indicating a stronger scatter of skewness values.

The concentration, or peakedness, of a probability distribution is quantified by the \emph{kurtosis}, or fourth standardized moment,

\begin{equation}
g_2=\frac{m_4}{m_2^2}-3=\frac{E\left[(X-\mu)^4\right]}{(E[(X-\mu)^2])^2}-3
\label{eq_kurt}
\end{equation}

\noindent
(e.g. \citealt{craw,press2007}). Here $m_4$ denotes the fourth sample moment about the mean, and $m_2$ is the second sample moment about the mean. If the probability distribution is Gaussian, the kurtosis is zero by construction. For each value of $\beta$, we calculate $g_2$ for all $N'$ realizations of flux distributions; we present the corresponding histograms in Fig. \ref{kurtosis}. For $\beta<1$, the average values of kurtosis are in agreement with zero (within errors), indicating agreement with Gaussian flux distributions. For $\beta\gtrsim1$, the mean values of kurtosis deviate significantly from zero; the mean values found in {\sc runs A and B} are in agreement within (standard) errors. Notably, the average kurtosis becomes negative, meaning the simulated flux distributions are, in average, platykurtic -- they are too flat compared to Gaussian profiles. This is indeed the signature we expect if a single, uni-modal distribution breaks up into multiple overlapping distributions.

In order to provide a more quantitative assessment, we check how much the averages of skewness and kurtosis deviate from zero in units of the standard error of the mean, which is given by

\begin{equation}
\sigma_{\overline x} = \frac{\sigma}{\sqrt{N'}}
\label{eq_stderr}
\end{equation}

\noindent
\citep{diek}. Here $N'$ is -- again -- the number of trials, being 100 for {\sc run A} and 500 for {\sc run B}. The results are shown in Tables \ref{tbl1} and \ref{tbl2}. Table \ref{tbl1} shows that the average values of skewness never deviate by more than $\approx2\,\sigma_{\overline x}$ from zero. In contrast, Table \ref{tbl2} shows that the average values of kurtosis deviate significantly -- meaning by more than $5\,\sigma_{\overline x}$ -- from zero for $\beta\gtrsim1.0$.

\begin{deluxetable}{lccccccccc}
\tablecolumns{10}
\tablewidth{0pc}
\tablecaption{{\sc Kurtosis} statistics for {\sc runs A and B}. Deviations from zero are in units of $\sigma_{\overline x}$, i.e. the standard error of mean. \label{tbl2}}
\tablehead{
\colhead{$\beta$} & 
\colhead{$0.0$} & \colhead{$0.25$} & \colhead{$0.5$} & \colhead{$0.75$} & \colhead{$1.0$} 
& \colhead{$1.25$} & \colhead{$1.5$} & \colhead{$1.75$} & \colhead{$2.0$}
}
\startdata
{\sc Run A:} & & & & & & & & & \\
mean & $-$0.004 & $-$0.003 & $-$0.002 & $-$0.009 & $-$0.059 & $-$0.184 & $-$0.345 & $-$0.494 & $-$0.612 \\
standard deviation & 0.052 & 0.052 & 0.053 & 0.072 & 0.154 & 0.301 & 0.417 & 0.496 & 0.560 \\
standard error & 0.005 & 0.005 & 0.005 & 0.007 & 0.015 & 0.030 & 0.042 & 0.050 & 0.056\\
deviation from 0 ($\sigma_{\overline x}$) & 0.710 & 0.579 & 0.375 & 1.222 & 3.833 & 6.097 & 8.272 & 9.960 & 10.92 \\
&&&&&&&&& \\
{\sc Run B:} & & & & & & & & & \\
mean & 0.003 & 0.001 & $-$0.003 & $-$0.021 & $-$0.085 & $-$0.217 & $-$0.381 & $-$0.525 & $-$0.635 \\
standard deviation & 0.114 & 0.114 & 0.114 & 0.131 & 0.203 & 0.323 & 0.435 & 0.518 & 0.580 \\
standard error & 0.005 & 0.005 & 0.005 & 0.006 & 0.009 & 0.014 & 0.019 & 0.023 & 0.026\\
deviation from 0 ($\sigma_{\overline x}$) & 0.588 & 0.196 & 0.588 & 3.585 & 9.363 & 15.02 & 19.59 & 22.66 & 24.48 \\
\enddata
\end{deluxetable}

\section{DISCUSSION}

\noindent
Our analysis provides a valuable test of the statistical emission properties of active galactic nuclei. We explore the relation between the shapes of the flux distributions (histograms) of red-noise lightcurves on the one hand, and the power-law indices of the temporal power spectra (periodograms) of those lightcurves on the other hand. For values $\beta\lesssim1$, the flux distributions are in agreement (see Tables \ref{tbl1}, \ref{tbl2}) with being Gaussian in average. This picture changes gradually for $\beta\gtrsim1$: with increasing values of $\beta$ the flux distributions tend to break up into multiple, overlapping sub-distributions (illustrated in Fig. \ref{chi_square}), leading to statistically significant deviations from Gaussian distributions (Fig. \ref{kurtosis}; Tables \ref{tbl1}, \ref{tbl2}). Turning this finding around, we may conclude: \emph{if} the power-law index is smaller than unity \emph{and} observations find multi-modal flux distribution \emph{then} the emission \emph{does not} originate from a single, uniform stochastic emission process.

Recent observations point out the necessity to analyze flux distributions carefully when addressing the emission statistics of AGN. \cite{dod} gave observational evidence for multiple emission states of Sgr A*. They concluded that its near-infrared flux distribution deviates systematically from realistic analytical profiles, especially log-normal distributions. This discrepancy can be resolved by assuming a superposition of two temporally distinct states of activity: a ``quiescent'' state at low energies, and a ``flare'' state at high energies.

\citet{trip} analyzed the high-frequency radio lightcurves of six AGN. Five of the six sources showed an unexpected pattern: whereas their periodograms followed red noise laws (with $0.4\lesssim\beta\lesssim0.7$)\footnote{When taking into account the measurement noise, the \emph{intrinsic} values of $\beta$ are slightly larger but still well below unity (see Sect.~5.3 of \citealt{trip} for details).}, their flux distributions showed strong bi- or even multi-modality. This behavior led \citet{trip} to suspect that the emission observed originates from distinct emission states. One possible explanation is provided by temporally distinct states of activity, like quiescent and, potentially multiple, flare states (cf. \citealt{dod}). Another explanation is the superposition of flux from several, causally disconnected emission regions -- like different shock zones in radio jets. \cite{trip} argued qualitatively that for red-noise power spectra with $\beta\lesssim1$, the observed flux distributions should be uni-modal \emph{if} the emission originates from the same stochastic process. However, this argument was never tested quantitatively; such a test is now provided by the present study.

Our statistical analysis indicates that, at least for moderate red-noise power-law indices of $\beta\lesssim1$, the underlying time series should be drawn from a uni-modal, approximately Gaussian, distribution. Evidently, the recent observations by \citet{trip} suggest that the actual picture is more complicated: even if the power spectra apparently correspond to uniform red-noise processes, the emission can arise from multiple components -- where these ``components'' may be distinct in time (temporally distinct states of activity) as well as in space (like different emission zones in jets). A combined analysis of temporal power spectra and flux distributions may be a valuable tool for studying AGN structures in general: as the analysis is based on lightcurves only, it probes the structure of sources without actually resolving them spatially.

Mathematically, red-noise power spectra are due to correlations between adjacent data points in the underlying time series \citep{pres}. The actual power-law index is given by the strength of correlation, ranging from no correlation at all -- white noise, $\beta=0$ -- up to correlation over infinite times -- random walk noise, $\beta=2$. However, a \emph{physical} explanation for the occurrence of red-noise time series in many physical and astrophysical processes is not yet known, and is subject to ongoing studies \citep{pres,kaulakys2009,kelly2011}. Accordingly, new insights and surprises regarding the statistical properties of AGN emission should be expected for the future.

\section{CONCLUSIONS}

\noindent
This study presents a test of the statistical properties of emission from active galactic nuclei. Using the algorithm introduced by \citet{tim}, we simulate lightcurves based on the assumption of uniform stochastic red-noise ($0\leq\beta\leq2$) emission processes. We compare the resulting flux distributions to those of recent mm/radio observations by \citet{trip}. Our work arrives at the following conclusions:

\begin{enumerate}

\item  For power-law indices $\beta\lesssim1$, the flux distributions (histograms) of our simulated lightcurves agree with noisy Gaussian profiles. For $\beta\gtrsim1$, the flux distributions remain symmetric (zero skewness) but become increasingly platykurtic (negative kurtosis) with increasing values of $\beta$. In those cases, the flux distributions break up into several overlapping sub-distributions, making the overall distributions multi-modal.

\item  From the behavior of the flux distributions as function of $\beta$, we also conclude: \emph{if} the power-law index is smaller than unity \emph{and} observations unveil multi-modal flux distributions, \emph{then} the emission \emph{does not} arise from a single, uniform stochastic emission process. This relation may be used to probe the structure of AGN without resolving the target sources spatially.

\item  Comparing our statistical analysis with the recent observation of multi-modal mm/radio flux distributions of five AGN by \cite{trip}, it becomes evident that \cite{trip} indeed observed emission from multiple emission states -- temporally distinct states of activity and/or spatially disconnected emission regions -- in their sources. In general, it appears that AGN emission can originate from multiple emission processes even if their simple red-noise power spectra suggest otherwise.

\end{enumerate}

A combined analysis of temporal power spectra and flux distributions of AGN can, a priori, be applied to a large variety of data sets available, like e.g. the radio flux monitoring by \citet{hov7,hov8}. Even though a priori simple, it appears that the power of such an analysis has largely been overlooked so far. Accordingly, we expect new insights from extended statistical analyses of AGN lightcurves in the future.

\acknowledgments{\noindent We acknowledge financial support from the Korea Astronomy and Space Science Institute (KASI) via Research Cooperation Grant 2012-1-600-90.}

\end{document}